%% file: main.tex
\documentclass[journal]{IEEEtran}
\usepackage{amsthm,amsmath,comment,bbm,amssymb,color,graphicx}
\usepackage{mathtools}
\usepackage{comment}
\usepackage{url}
\usepackage{caption}
\usepackage{subcaption}
\usepackage{epstopdf}
\usepackage{tikz}
\usetikzlibrary{positioning,shapes,shadings,backgrounds,calc}
\usepackage{multirow}
\usepackage{multicol}
\usepackage{enumitem}
\usepackage{booktabs,makecell}
\usepackage{xcolor}

\usepackage{fancyhdr} 
\fancypagestyle{plain}{ 
  \fancyhf{} 
  \fancyhead[C]{\scriptsize This article has been accepted for publication in IEEE Transactions on Communications. This is the author's version which has not been fully edited and 
  
  content may change prior to final publication. Citation information: DOI 10.1109/TCOMM.2026.3668158} 
  \fancyfoot[C]{\scriptsize \copyright{} 2026 IEEE. All rights reserved, including rights for text and data mining and training of artificial intelligence and similar technologies. Personal use is permitted, 
  
  but republication/redistribution requires IEEE permission. See https://www.ieee.org/publications/rights/index.html for more information.}
}

\title{Faster-than-Nyquist Signaling in the Finite Time-Bandwidth Product Regime}

\begin{document}
\include{notation_new}

\author{Yong Jin Daniel Kim
\thanks{Y. J. D. Kim is with the Department of Electrical and Computer Engineering, Rose-Hulman Institute of Technology, Terre Haute, IN 47803, USA.}
}
\maketitle
\thispagestyle{plain} 

\begin{abstract}

This paper analyzes faster-than-Nyquist (FTN) signaling within a consistent framework based on a fixed time-bandwidth product (TBP), resolving potential ambiguities present in finite blocklength analysis. A key feature of FTN is its ability to increase the number of transmitted symbols in a given time and frequency resource, which can lower the rate penalties inherent in short packet communications. We derive tight bounds on the maximum channel coding rate (MCCR) and demonstrate that FTN's rate gains over Nyquist signaling can be higher in the finite TBP regime than in the asymptotic case. Performance is benchmarked against the theoretical optimum of transmitting prolate spheroidal wave functions, showing that a well-designed FTN system can closely approach this limit. We present practical design criteria, including the optimal time-acceleration factor for maximizing signaling dimensions, an optimized pulse shape that meets strict out-of-band constraints, and a turbo-equalization-based coding scheme that performs near the derived MCCR bounds. These findings establish FTN as a practical and near-optimal technique for enhancing the rate and reliability of latency-constrained communications.

\end{abstract}


\bstctlcite{IEEEexample:BSTcontrol}

\section{Introduction}

\IEEEPARstart{F}{aster}-than-Nyquist (FTN) signaling is a non-orthogonal transmission technique with a signaling rate exceeding the Nyquist limit for orthogonality \cite{anderson2013faster, fan2017faster, ishihara2021evolution}. This method can recover the spectral efficiency typically lost due to the excess bandwidth of practical pulse-shaping \cite{rusek2009constrained}. The resulting non-orthogonality introduces a controlled-amount of inter-symbol interference (ISI) at the receiver, which may be effectively compensated for using precoding and/or equalization techniques (see e.g., \cite{liveris2003exploiting, prlja2012reduced, sugiura2015frequency, kim2016faster}).  

There is a growing interest in designing (and optimizing) systems for the finite blocklength (FBL) regime, driven by the emergence of delay sensitive applications that demand high reliability \cite{shirvanimoghaddam2018short}. Communication in the FBL regime, where packet sizes are a few hundred coded bits or less, presents unique challenges. The block error rates (BLER) can no longer vanish for most channels of interest, and for a nonzero target BLER, there usually exists a rate penalty compared to the channel capacity. Furthermore, practical pulse shapes such as the root-raised cosine (RRC) with very small roll-off factors are difficult to implement due to their slow time decay, and even the ideal sinc pulses with Nyquist rate signaling are no longer optimal when sampling is restricted to a finite time window.

A unique characteristic of FTN is its ability to increase the blocklength by packing more symbols into a given time and frequency resource. This naturally raises the question of whether FTN can mitigate the aforementioned challenges of the FBL regime. Prior work, leveraging finite-blocklength information theory, confirmed that FTN can indeed lower the rate penalty associated with short packet transmissions \cite{mohammadkarimi2020channel}, \cite{zhang2025maximum}. However, these analyses were conducted for a fixed blocklength, which can create ambiguity, as FTN allows for sending more symbols in the same time duration or the same number of symbols in a shorter time, making blocklength a variable parameter. Furthermore, continuous-time parameters such as pulse-width and out-of-band (or out-of-interval) energy need to be taken into account in the analysis. 

This work addresses this gap by analyzing the maximum channel coding rate (MCCR) of FTN within a consistent framework based on a fixed time-bandwidth product (TBP). We define a signal as having a finite TBP if it is either time-limited with out-of-band (OOB) power below a tolerance level, or band-limited with out-of-interval (OOI) power below a tolerance level. The first model reflects the physical reality that transmitted signals are time-limited and have non-zero out-of-band energy, and the second model allows analysis of bandlimited pulse shapes, providing useful insights.\footnotemark 

\footnotetext{Several other continuous-time channel models exist \cite{wyner1966capacity}, \cite{gallager1968information}. A notable example is the model by Gallager \cite[Ch. 8]{gallager1968information}, where the signal is first constrained in time and then constrained in frequency by being sent through a perfectly bandlimited filter before transmission. Expanding on this model, \cite{jaffal2020time} derived MCCR in the finite TBP regime using prolate spheroidal wave functions as basis functions. While this approach also provides a consistent framework, particularly in the asymptotic TBP regime, the resulting signal remains strictly bandlimited. Our model, in contrast, captures the more realistic scenario of time-limited signals with out-of-band energy, though the results depend on the chosen tolerance level.}

Within this framework, we derive tight bounds on the MCCR for a fixed TBP and demonstrate that FTN's rate gains over Nyquist signaling can be more significant in the finite TBP regime than in the asymptotic case. In addition, with appropriately chosen pulse shapes FTN can perform close to the theoretical benchmark established by transmitting prolate spheroidal wave functions (PSWFs) as the basis functions. FTN can also achieve lower BLER for fixed coding rates, highlighting its potential to enhance the performance and reliability of short packet communications. 

We also present several design criteria for FTN systems operating in the finite TBP regime. We derive the time acceleration factor of FTN that yields the maximum signaling dimensions for a given TBP and show it is strictly below the asymptotic optimum limit, i.e., below $\frac{1}{1+\beta}$, where $\beta$ is the roll-off factor of the utilized pulse shape. This demonstrates that optimal system design in the finite TBP regime deviates fundamentally from principles derived from the asymptotic analysis. Additionally, we propose an optimized pulse shape that meets a strict OOB constraint and provide guidelines on pulse design in the finite TBP regime. Finally, we demonstrate that a practical FTN coding scheme based on turbo-equalization can achieve performance near MCCR limits. 

The remainder of this paper is organized as follows. Section II details the FTN channel model and assumptions. Section III develops the capacity and the MCCR bounds for FTN signaling. Section IV presents numerical results and discussions on the merits of FTN. Section V covers system design considerations, including optimal signaling rates, pulse shaping, and coding. Section VI provides a summary.

\section{Discrete-time FTN Channel Model}
This section formulates a discrete-time model for the FTN channel, representing it as a set of $N$ parallel Gaussian channels. We begin by formally defining the time-bandwidth product.

\subsection{Time-bandwdith product}

Consider a data communication over a duration of $T_x$ seconds and a bandwidth of $W$ Hz. The time-bandwidth product (TBP) is a dimensionless quantity defined as:
    \begin{equation} \label{eq:TBP}
        \Omega\triangleq2WT_x. 
    \end{equation}
In this work, a signal $x(t)$ is considered to have a TBP $\Omega$ under one of two conditions:

\begin{enumerate}[leftmargin=*]
    \item Time-limited with an out-of-band (OOB) constraint: The signal $x(t)$ is strictly time-limited to $T_x$ seconds, and its OOB power is less than a fraction $\epsilon_W>0$ of its total average power $P$. This is expressed as
    \begin{equation} \label{eq:OOB} \begin{split}
        P - \mathbb{E} \left[  \frac{1}{T_x} \int_{-W}^W \lvert \hat{x}(f) \rvert ^2 df \right] \leq \epsilon_W P,
    \end{split} \end{equation}
    where $\hat{x}(f)$ is the Fourier transform of $x(t)$.

    \item Band-limited with an out-of-interval (OOI) constraint: The signal $x(t)$ is strictly band-limited to $W$ Hz, and its OOI power is less than a fraction $\epsilon_T>0$ of its total average power. This is expressed as
    \begin{equation} \label{eq:OOI} \begin{split} 
        P - \mathbb{E} \left[ \frac{1}{T_x} \int_{\mathcal{T}_x} \lvert x(t) \rvert ^2 dt \right] \leq \epsilon_T P, 
    \end{split} \end{equation}
    \noindent where $\mathcal{T}_x$ is the time interval of duration $T_x$ seconds in which the signal's energy is concentrated.
\end{enumerate}

\subsection{FTN system model and assumptions}

Let $p(t)$ be a real-valued, unit energy pulse with a Nyquist symbol time of $T$ seconds. In an FTN system, symbols are transmitted at an accelerated rate of $\frac{1}{\tau T}$ symbols per second, where $\tau<1$ is the time-acceleration factor. The complex baseband FTN signal $x(t)$ is given by
    \begin{equation} \label{eq:x(t)}
        x(t)=\sqrt{\frac{PT_x}{N}}\sum_{n=0}^{N-1}{x_np(t-n\tau T)}, 
    \end{equation}
where $\{x_n\}$ are the complex data symbols and $P$ is the average signal power:
    \begin{equation} \label{eq:P}
        P=\mathbb{E}\left[ \frac{1}{T_x} \int_{-\infty}^\infty\lvert x(t) \rvert^2 dt \right].
    \end{equation}
We assume that the data symbols are uncorrelated and identically distributed (\textit{u.i.d.}) with zero mean and unit variance. Under this assumption, the power spectral density (PSD) of the FTN signal is $P|\hat{p}(f)|^2$ and independent of $\tau$.\footnote{Higher MCCR may be obtained by precoding or allocating power non-uniformly across the symbols \cite{mohammadkarimi2020channel}, \cite{ishihara2021eigendecomposition}, but these necessarily alter the shape of PSD and can result in bandwidth expansion if not designed carefully \cite{kim2010spectrum}.}

The total time duration $T_x$ required to transmit $N$ symbols depends on $\tau$ and the pulse-width of $p(t)$, $T_p$, such that $T_x = (N-1)\tau T+T_p$. Therefore, for a time-limited pulse $p(t)$, FTN can send
    \begin{equation} \label{eq:N}
        N=\left\lfloor\frac{T_x-T_p}{\tau T}+1\right\rfloor = \left\lfloor \frac{\Omega-c}{\tau(1+\beta)} + 1 \right\rfloor
    \end{equation}
number of symbols within a TBP of $\Omega$, where the last expression is obtained by setting $T=\frac{1+\beta}{2W}$ and denoting $c\triangleq2WT_p$ as the TBP of the pulse. We note that FTN can increase $N$ by lowering $\tau$ for a fixed TBP, making the blocklength a variable parameter. With the \textit{u.i.d.} data symbols, the OOB constraint also simplifies to a condition only on the pulse:
    \begin{equation} \label{eq:OOB_FTN} 
        1-\int_{-W}^W\lvert \hat{p}(f) \rvert ^2 df\leq\epsilon_W,
    \end{equation}
where $\hat{p}(f)$ is the Fourier transform of $p(t)$. In other words, the OOB constraint is met if the time-limited pulse $p(t)$ has fraction of energy outside the band $(-W,W)$ at most $\epsilon_W$. For a bandlimited pulse, the OOI constraint is expressed as 
    \begin{equation} \label{eq:OOI_FTN} 
        1-\frac{1}{N}\sum_{n=0}^{N-1}\int_{\mathcal{T}_x}\lvert p(t-n\tau T) \rvert ^2 dt\leq\epsilon_T,
    \end{equation}
which depends on $N$ and $\tau$, and thus the maximum blocklength is determined numerically by computing for the maximum $N$ while \eqref{eq:OOI_FTN} is satisfied. 

The signal is transmitted over a complex additive white Gaussian noise (AWGN) channel with a noise PSD of $N_0$. Without loss of generality, we assume a receiver employing a filter matched to $p(t)$ and sampling at the FTN rate of $\frac{1}{\tau T}$.\footnote{The matched filter receiver yields sufficient statistics, as the set of signals $\{p(t-n\tau T)\}_{n=0}^{N-1}$ is linearly independent and spans the signal space.} The resulting discrete-time vector of matched filter outputs can be expressed as:

    \begin{equation} \label{eq:y}
        \mathbf{y}=\sqrt{\frac{PT_x/N}{N_0}}\mathbf{Hx}+\mathbf{z},
    \end{equation}
where $\mathbf{y}$, $\mathbf{x}$, and $\mathbf{z}$ are $N\times 1$ vectors representing the outputs, data symbols, and colored noise samples with the covariance of $\text{cov}(\mathbf{z})=\mathbf{H}$, respectively. The $N \times N$ matrix $\mathbf{H}$ is a symmetric Toeplitz matrix:
    \begin{equation} \label{eq:H}
        \mathbf{H}=
        \begin{bmatrix}
            h_0     & h_{-1}    & \cdots    & h_{-(N-1)} \\
            h_1     & h_0       & \cdots    & h_{-(N-2)} \\
            \vdots  & \vdots    & \ddots    & \vdots \\
            h_{N-1} & h_{N-2}   & \cdots    & h_0       
        \end{bmatrix},
    \end{equation}
where $\{h_n\}$ are the samples of the pulse autocorrelation, $h_n=\int_{-\infty}^{\infty}p(t)p(t-n\tau T)dt$. This channel matrix is full rank for finite $N$ when $p(t)$ is either time-limited or band-limited \cite{kim2016properties}, \cite{gattami2015time}.

\subsection{$N$-parallel Gaussian channel formulation for FTN}

\begin{figure}
    \centering
    \input{figNparallChan}
    \caption{$N$-parallel Gaussian channel formulation of FTN signaling, where the noise in the $n$-th channel has variance $\sigma_n^2 = (\rho\frac{\Omega}{N}\lambda_n)^{-1}$. The SNR of the $n$-th channel is $1/\sigma_n^2$.}
    \label{fig:NparallChan}
\end{figure}

The coupled FTN channel  may be diagonalized into a set of parallel, independent Gaussian channels via an eigen-decomposition on the channel matrix, $\mathbf{H} = \mathbf{U\Lambda U}^T$, where the columns of the orthogonal matrix $\mathbf{U}$ are the eigenvectors, $\mathbf{U}^T$ is the matrix transpose of $\mathbf{U}$, and $\mathbf\Lambda$ is a diagonal matrix with the eigenvalues, $\{\lambda_n\}$, on the diagonal. Applying this transformation and normalizing the result yields the canonical $N$-parallel channel model shown in Fig. \ref{fig:NparallChan}:
    \begin{equation} \label{eq:y_tilde}
        \boxed{\tilde{\mathbf{y}} = \tilde{\mathbf{x}}+\tilde{\mathbf{z}}.}
    \end{equation}
Here, $\tilde{\mathbf{x}} \triangleq \mathbf{U}^T \mathbf{x}$ is the transformed input vector, which remains \textit{u.i.d.} with unit variance due to $\text{cov}(\tilde{\mathbf{x}})=\text{cov}(\mathbf{x})$.\footnote{The \textit{u.i.d.} assumption, $\text{cov}(\tilde{\mathbf{x}})=\mathbf{I}$, means that the input must satisfy the average power constraint, $\mathbb{E}\{\lvert \tilde{x}_n \rvert ^2\}=1$ for all $n$. In the MCCR analysis, sometimes it will be convenient to impose a stricter condition of constant input power for all realizations, i.e., $\lvert \tilde{x}_n \rvert^2=1$ for all $n$. The constant power implies that $\frac{1}{N} \|\mathbf{x}\|^2=1$ (i.e., equal power for individual codewords) due to $\|\mathbf{x}\|=\|\tilde{\mathbf{x}}\|$. We show in section \ref{subsection:MC} that the inputs $\{\tilde{x}_n\}$ under the constant power constraint can achieve the channel capacity as TBP tends to infinity (see also \cite{scarlett2016dispersion} for asymptotic optimality of constant input power). The type of constraint used will be stated explicitly in the MCCR analysis.} The vector $\tilde{\mathbf{z}} \triangleq \sqrt{\frac{N_0}{PT_x/N}} \mathbf{\Lambda}^{-1} \mathbf{U}^T \mathbf{z}$ contains independent Gaussian noise samples, where the noise variance for the $n$-th subchannel is $\sigma_n^2 = (\rho \frac{\Omega}{N} \lambda_n)^{-1}$, and $\rho \triangleq \frac{P}{2WN_0}$ is the signal-to-noise ratio (SNR). The $n$-th subchannel therefore has the signal-to-noise ratio of $\text{SNR}_n=1/\sigma_n^2=\rho \frac{\Omega}{N} \lambda_n$. We note that FTN can increase the number of parallel channels by lowering $\tau$ in a fixed TBP.

As evident from the $N$-parallel channel model formulation, the eigenvalues $\{\lambda_n\}$ of the channel matrix $\mathbf{H}$ determine the quality of each subchannel. These eigenvalues are known to be strictly positive, add to $N$, and well approximated by samples of the folded spectrum for large $N$ \cite{kim2016properties}: 
    \begin{equation} \label{eq:lambda_n_approx}
        \lambda_n \approx (\tau T)^{-1} \hat{p}_\text{folded}(f_n) \text{ for } f_n=\tfrac{n}{N\tau T},
    \end{equation}
where the folded-spectrum is defined by $\hat{p}_\text{folded}(f) \triangleq \sum_{k=-\infty}^{\infty} \left\vert \hat{p}\left(f-\frac{k}{\tau T}\right) \right\vert ^2$. Closed-form expressions for the folded-spectrum are available for many pulses of interest \cite{kim2016properties}. 

\begin{figure}
    \centering
    \includegraphics[width=\linewidth]{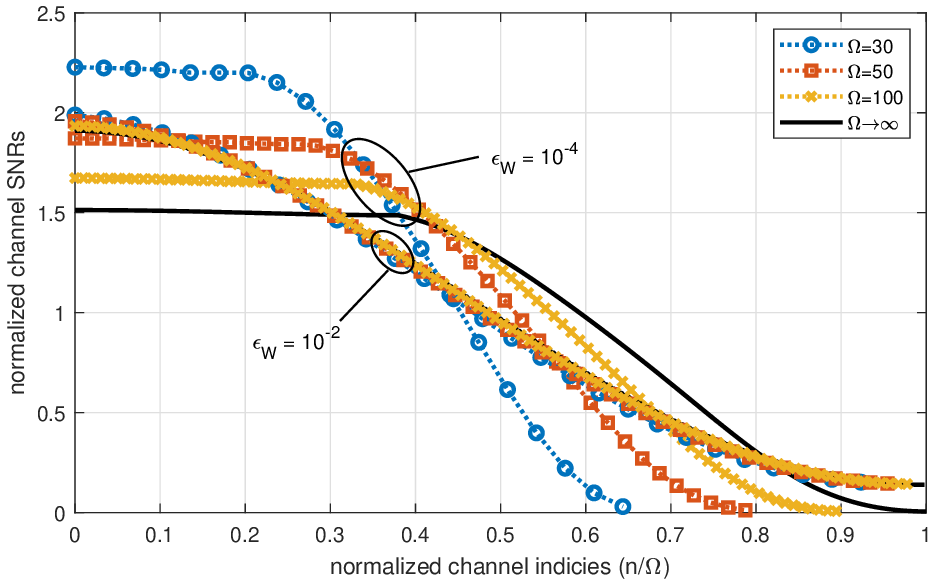}
    \caption{Normalized SNRs of the parallel channels ($\sigma_n^{-2}/\rho$) over normalized channel indices $n/\Omega$ for $n=0,1,\cdots,N-1$ for varying OOB levels and TBPs. RRC pulses with roll-off $\beta=0.5$, time-truncated so that OOB energy satisfies $\epsilon_W=10^{-2}$ or $10^{-4}$, and $\tau=1/(1+\beta)$ considered.}
    \label{fig:channelSNR_normindicies}
\end{figure}

Fig. \ref{fig:channelSNR_normindicies} plots the normalized SNRs of the $N$-parallel channels (i.e., $\sigma_n^{-2}/\rho = \tfrac{\Omega}{N}\lambda_n$) for varying OOB levels and TBPs. These are plotted over normalized channel indices $\frac{n}{\Omega}$ for $n=0,1,\cdots,N-1$. The eigenvalues $\{\lambda_n\}$ are computed numerically, while the folded-spectrum \eqref{eq:lambda_n_approx} is used for $\Omega\to\infty$. We observe that as TBP increases the normalized SNRs converge towards the asymptote \eqref{eq:lambda_n_approx}, but the convergence is slow for more stringent OOB levels (e.g., $\epsilon_W=10^{-4}$). These imply that while \eqref{eq:lambda_n_approx} yields an excellent approximation in many cases, its accuracy is not guaranteed when TBP is low, which is the focus of this work. Deriving exact analytical eigenvalues of $H$ in the finite TBP regime is a challenging mathematical problem. For these reasons, all results presented in this work are based on the direct numerical computation of the eigenvalues.

\section{MCCR of FTN in the Finite TBP Regime} \label{section:MCCR}

This section presents the channel capacity and MCCR of FTN, including a normal approximation, a converse upper-bound, and an achievability lower-bound, in the finite TBP regime. 

\subsection{Capacity of FTN signaling}

The channel capacity is the maximum rate at which information can be communicated with vanishing probability of error in the asymptotic TBP regime. The channel capacity of FTN signaling may be obtained from the FTN channel model \eqref{eq:y_tilde}. First, capacity of $N$-parallel Gaussian channel with SNR of the $n$-th channel equal to $1/\sigma_n^2$ is given by \cite{cover1999elements}
    $$ C=\lim_{N\to\infty} \frac{1}{\Omega} \sum_{n=0}^{N-1} \log_2\left(1+\frac{1}{\sigma_n^2}\right) \text{ [bps/Hz]}.$$
Noting that for FTN, $\sigma_n^2=(\rho\tfrac{\Omega}{N}\lambda_n)^{-1}$ and $T_x \approx N\tau T$ for large TBP, and using the asymptotic distribution of the eigenvalues \eqref{eq:lambda_n_approx}, the capacity converges in the limit as \cite{rusek2009constrained, kim2016properties}
    \begin{equation} \label{eq:C_FTN}
        \boxed{ 
        C_{\text{FTN}} = \frac{1}{2W} \int_{-\frac{1}{2\tau T}}^{\frac{1}{2\tau T}}  \log_2(1+\rho2W \hat{p}_{\text{folded}}(f))df. }
    \end{equation}
The above expression holds for any $\tau \leq 1$ and is non-decreasing with the signaling rate of FTN. In the limit $\tau\to0$, the folded-spectrum converge as $\hat{p}_{\text{folded}}(f) \to \lvert \hat{p}(f)\rvert^2$, and the capacity becomes: $\frac{1}{2W} \int_{-\infty}^{\infty} \log_2(1+\rho2W \lvert\hat{p}(f)\rvert^2)df$. When $p(t)$ is bandlimited, the capacity reaches maximum with $\tau = \tau_0 \triangleq \tfrac{1}{1+\beta}$, where $\beta$ denotes the roll-off factor of the modulating pulse, with the capacity given by $\frac{1}{2W} \int_{-W}^{W} \log_2(1+\rho2W \lvert\hat{p}(f)\rvert^2)df$.

\subsection{Normal approximation}

A normal approximation of MCCR is first derived below from the $N$-parallel Gaussian channel model of FTN.

\begin{prop}[Normal approximation for FTN] \label{prop:FTN_NA}
    Assume the constant power constraint, $\lvert \tilde{x}_n \rvert^2=1$ for all $n$, and a fixed $\tau<1$. A normal approximation (NA) of MCCR [bps/Hz] for the $N$-parallel Gaussian channel model of FTN at TBP $\Omega$ with average BLER $P_e$ is given by 
    \begin{equation} \label{eq:R_NA}
        \boxed{ R_{\text{NA}} = C_{\text{NA}} - \sqrt{\frac{V_{\text{NA}}}{\Omega}}\log_2(e) Q^{-1}(P_e) +\frac{\log_2(\Omega)}{2\Omega}, }
    \end{equation}
where $Q^{-1}$ denotes the inverse $Q$-function (i.e., inverse function of the complementary Gaussian cumulative distribution function (CDF)) and
    \begin{equation*}
    \begin{split}
    &C_\text{NA} \triangleq \frac{1}{\Omega} \sum_{n=0}^{N-1} \log_2 \left( 1+\frac{1}{\sigma_n^2} \right), \\
    &V_{\text{NA}} \triangleq \frac{1}{\Omega} \sum_{n=0}^{N-1} \left( 1-\frac{1}{(1+\frac{1}{\sigma_n^2})^2}\right), 
    \end{split}
    \end{equation*}
are the finite blocklength approximations to the channel capacity and the channel dispersion, respectively, where $\sigma_n^2 = (\rho \tfrac{\Omega}{N} \lambda_n )^{-1}$ and $N$ given by \eqref{eq:N}. The NA may also be stated in terms of BLER $P_e$ as
    \begin{equation} \label{eq:P_e}
        \boxed{ P_e \approx Q \left( 
        \frac{C_{\text{NA}} - R_{\text{NA}} + \frac{\log_2(\Omega)}{2\Omega}} {\sqrt{\frac{V_{\text{NA}}}{\Omega}}\log_2(e)} \right).}
    \end{equation}

\end{prop}

Proof of the proposition is given in Appendix \ref{Appendix:asym_expansion_MC} and is based on asymptotic expansion of the converse bound presented in section \ref{subsection:MC}. Here, in lieu of the complete proof, we present a simpler (but partial) proof of the proposition that recovers $R_\text{NA}$ up to the second-order term\footnote{This proof approach, however, yields an incorrect third-order term that is unbounded with decreasing $\tau$. See Appendix \ref{Appendix:asym_expansion_MC} for the complete proof.}. First consider real, $N$-parallel AWGN channels with $m$ real-valued symbols sent in each channel. The $n$-th channel output has the form $\mathbf{r}_n=\mathbf{a}_n+\mathbf{w}_n$, where each vector has length $m$, the channel input $\mathbf{a}_n$ is assumed to have fixed power, i.e., $\frac{1}{m} \|\mathbf{a}_n\|^2=P_n$, and the additive noise $\mathbf{w}_n \sim \mathcal{N}(\mathbf{0},\omega_n^2\mathbf{I})$ is independent of $\mathbf{a}_n$. The $n$-th channel has $\text{SNR}_n=\frac{P_n}{\omega_n^2}$. A normal approximation of this $N$ real-valued parallel Gaussian channel in bits per channel use is given by \cite[Theorem 10]{erseghe2016coding}
    $$R_N = C_N-\sqrt{\frac{V_N}{mN}}\log_2(e)Q^{-1}(P_e) + \frac{\log_2{mN}}{2mN},$$
where $C_N$ and $V_N$ are
    \begin{equation*}
    \begin{split}
        &C_N=\frac{1}{N}\sum_{n=0}^{N-1} \frac{1}{2} \log_2(1+\text{SNR}_n),\\
        &V_N=\frac{1}{N} \sum_{n=0}^{N-1} \frac{1}{2} \left(1-\frac{1}{(1+\text{SNR}_n)^2} \right).
    \end{split}
    \end{equation*}
Now, coding over the complex channels may be seen as coding over real channels using a blocklength of $2N$ \cite{gursoy2013throughput}. First, express the FTN channel output \eqref{eq:y_tilde} as 
    $$\tilde{\mathbf{y}} = \tilde{\mathbf{y}}_{\text{Re}}+j\tilde{\mathbf{y}}_{\text{Im}}=(\tilde{\mathbf{x}}_{\text{Re}}+\tilde{\mathbf{z}}_{\text{Re}})+j(\tilde{\mathbf{x}}_{\text{Im}}+\tilde{\mathbf{z}}_{\text{Im}}),$$
where the subscripts Re and Im denote the real and imaginary components, respectively, of the corresponding vectors. This is rearranged into $[\tilde{\mathbf{y}}_{\text{Re}},\tilde{\mathbf{y}}_{\text{Im}}] =[\tilde{\mathbf{x}}_{\text{Re}},\tilde{\mathbf{x}}_{\text{Im}}] + [\tilde{\mathbf{z}}_{\text{Re}},\tilde{\mathbf{z}}_{\text{Im}}],$
which is an $N\times2$ matrix formed by concatenating $\tilde{\mathbf{y}}_{\text{Re}}$ and $\tilde{\mathbf{y}}_{\text{Im}}$. We may treat the two symbols in each row being transmitted in one of the $N$-parallel channels. The real and imaginary noise components, $\tilde{\mathbf{z}}_{\text{Re}}$ and $\tilde{\mathbf{z}}_{\text{Im}}$, are independent due to the circular symmetry of the additive complex Gaussian noise and the variances of the individual entries are halved from that of the complex noise counterpart $\tilde{\mathbf{z}}$. Setting $m=2$ to account for two real-valued symbols per complex channel, $\text{SNR}_n=\sigma_n^{-2}$ (due to $P_n=0.5$ for the constant input power and $\omega_n^2=\sigma_n^2/2$) for FTN, doubling the rates by noting that two (real-valued) symbols are sent for every one complex symbol, and changing the units to bps/Hz by noting that $N$ FTN symbols are sent in $T_x$ seconds and $2W$ Hz, we obtain the expression \eqref{eq:R_NA} up to the second-order term. 

\begin{nrem}
    The unit of $R_\text{NA}$ may be converted to [bps] by multiplying by $2W$, or to [bpcu] by multiplying by $\frac{\Omega}{N}$. 
\end{nrem}

\begin{nrem}
    For real-valued FTN channels, $C_{\text{NA}}$ and $V_{\text{NA}}$ should be replaced by $\frac{1}{2} C_{\text{NA}}$ and $\frac{1}{2} V_{\text{NA}}$, respectively (due to the information density of real-valued channels being half of that of complex-valued channels), and SNR $\rho$ is replaced by $\rho_{\text{real}} \triangleq \frac{P}{WN_0}$. The third order term is the same. 
\end{nrem}

\begin{nrem}
    As $\Omega\to\infty$ for any fixed $\tau<1$, we have $C_{\text{NA}} \to C_{\text{FTN}}$ and $R_{\text{NA}} \to C_{\text{NA}}$, thus NA is asymptotically tight. Similarly, using the asymptotic eigenvalue distribution 
    \eqref{eq:lambda_n_approx}, we have
        $$V_{\text{NA}} \to \frac{1}{2W} \int_{-\frac{1}{2\tau T}}^{\frac{1}{2\tau T}} \left(1-\frac{1}{(1+\rho2W\hat{p}_{\text{folded}}(f))^2}\right)df,$$
    which agrees with the channel dispersion term of \textit{i.i.d.} FTN derived in \cite{mohammadkarimi2020channel} when $\tau=\tau_0$ and data symbols are real-valued.
\end{nrem}

In the literature, NA is known to yield a simple, yet remarkably tight, approximation to MCCR (or BLER) in AWGN channels \cite{polyanskiy2010channel}. However, it may be inaccurate for $\Omega$ very small ($\lesssim 100$) or when the selected rate is significantly smaller than the capacity. For these reasons, converse and achievability bounds which provide true upper and lower bounds, respectively, of MCCR are often necessary. These bounds are investigated next. 

\subsection{Converse upper-bound} \label{subsection:MC}

We consider the Polyanskyi-Poor-Verd\'u (PPV) meta-converse (MC) bound \cite{polyanskiy2010channel}, which is one of the tightest upper-bounds available for AWGN channel. The MC bound is based on a binary hypothesis test of given observation pair $(\mathbf{x},\mathbf{y})$ being distributed as either
    $$\text{H}_1:(\mathbf{x},\mathbf{y}) \sim p_{y|x}p_x \text{ or } 
    \text{H}_0:(\mathbf{x},\mathbf{y}) \sim q_y p_x$$
for some choice of the probability density function (PDF) $q_y$. The Neyman-Pearson test is a log-likelihood ratio test given by
    $$\Lambda(\mathbf{x},\mathbf{y}) = \frac{1}{N} \ln \frac{p_{y|x}(\mathbf{y}|\mathbf{x})}{q_y(\mathbf{y})} \overset{\text{H}_0}{\underset{\text{H}_1}\lessgtr} \lambda,$$
for some $\lambda$. The two associated probability of errors (i.e., the missed detection (MD) and the false alarm (FA) probabilities) are, for a given $\mathbf{x}$,
    \begin{equation*} \begin{split}
        &P_{\text{MD}}(\mathbf{x},\lambda) = P\left[\Lambda(\mathbf{x},\mathbf{y}) < \lambda | \text{H}_1,\mathbf{x}\right] \text{ and}\\
        &P_{\text{FA}}(\mathbf{x},\lambda) = P\left[\Lambda(\mathbf{x},\mathbf{y}) > \lambda | \text{H}_0,\mathbf{x}\right],
    \end{split} \end{equation*}
and for a given codebook $\mathcal{C}$,
    \begin{equation*} \begin{split}
        &P_{\text{MD}}(\lambda) = \sum_{\mathbf{x}\in\mathcal{C}} P_{\text{MD}}(\mathbf{x},\lambda) p_x(\mathbf{x}),\\
        &P_{\text{FA}}(\lambda) = \sum_{\mathbf{x}\in\mathcal{C}} P_{\text{FA}}(\mathbf{x},\lambda) p_x(\mathbf{x}).
    \end{split} \end{equation*}
Using the notations above, one version of MC can be stated as follows (see \cite[Theorem 27]{polyanskiy2010channel} for more general version of the theorem).

    \begin{thm}[MC \cite{polyanskiy2010channel}, \cite{erseghe2016coding}] \label{thm:PPV_MC}
        Assume that $p_x(\mathbf{x}) = \frac{1}{M}$ (equally-likely message) for all $\mathbf{x}\in\mathcal{C}$ and $P_e$ is the average BLER. If $q_y$ is chosen in such a way that both MD and FA probabilities are independent of $\mathbf{x}$, then for a fixed $P_e$, the code rate $R$ is upper-bounded as
            $$R \leq -\frac{1}{N} \log_2{P_{\text{FA}}(\lambda)} \text{ [bpcu]},$$
        where $\lambda$ satisfies $P_{\text{MD}}(\lambda)=P_e$.
    \end{thm}
    \begin{proof}
        See \cite[Theorem 28]{polyanskiy2010channel}.
    \end{proof}
 
The MC bound for the considered FTN signaling is given in the following theorem.

    \begin{thm}[MC for FTN] \label{thm:MC_FTN}
        Let $\mathcal{X}^2(k,\nu)$ denote the complex noncentral chi-square distribution with degree of freedom $k$ and noncentrality parameter $\nu$. Let $U_n \sim \mathcal{X}^2(1,1+\sigma_n^2)$ and $V_n \sim \mathcal{X}^2(1,\sigma_n^2)$ be two random variables for $n=0,1,\dots,N-1$, both independent in $n$. The MC bound for the FTN channel model \eqref{eq:y_tilde} under the constant power constraint, $|\tilde{x}_n|^2=1, \forall n$, is given by
	\begin{equation} \label{eq:MC_FTN}
            \boxed{ R \leq -\frac{1}{\Omega} \log_2{P\left[ \frac{1}{N} \sum_{n=0}^{N-1} \frac{1}{\sigma_n^2} U_n < \lambda \right]}}  \text{ [bps/Hz]},
        \end{equation}
        
        \noindent where $\lambda$ is chosen to satisfy
        \begin{equation} \label{eq:Pe_MC}
	   P_e=P\left[\frac{1}{N} \sum_{n=0}^{N-1} \frac{1}{1+\sigma_n^2} V_n > \lambda\right].
        \end{equation}
    \end{thm}
    \begin{proof}
        See Appendix \ref{Appendix:ProofMC}.
    \end{proof}

\begin{nrem}
    In Appendix \ref{Appendix:asym_expansion_MC}, an asymptotic expansion of the MC bound as $\Omega\to\infty$ is shown to recover $R_{\text{NA}}$ \eqref{eq:R_NA}, thus completing the proof of Proposition \ref{prop:FTN_NA}. 
\end{nrem}

Evaluating the MC bound directly is difficult, because PDF of the weighted sum of $\mathcal{X}^2(1,\mu)$ is unavailable in a closed form and the tail probabilities can get smaller than typical computing precision limit, thus making the Monte-Carlo methods infeasible. Fortunately, one may accurately approximate these bounds using methods such as the saddlepoint approximations (e.g., \cite{font2018saddlepoint}, \cite{maya2016closed}). Appendix \ref{Appendix:Approx} details an approximation of the MC bound using the method in \cite{maya2016closed}.

\subsection{Achievability lower-bound}

The random-coding union (RCU) achievability bound \cite{polyanskiy2010channel} is based on Shannon’s random coding argument. 

    \begin{thm}[RCU \cite{polyanskiy2010channel}, \cite{erseghe2016coding}] \label{thm:RCU}
        Let a $(M,N)$ code for a channel $p_{\mathbf{y}\mid\mathbf{x}}$ consists of $M$ codewords of length $N$ with the rate $R=\frac{1}{N} \log_2M$ [bpcu]. Then for every $M$ and $N$, there exists a code whose error probability is upper-bounded by
            $$P_e \leq \mathbb{E}_{\mathbf{x},\mathbf{y}} [\min\left\{1,(M-1)g(\mathbf{x},\mathbf{y})\right\} ]$$
        with $\mathbf{x} \sim p_\mathbf{x}$, $\mathbf{y} \sim p_{\mathbf{y}\mid\mathbf{x}}$, where $g(\mathbf{x},\mathbf{y})$ is the pairwise error probability, defined as
            $$g(\mathbf{x},\mathbf{y}) \triangleq P\left[p_{\mathbf{y}|\mathbf{x}}(\mathbf{y}|\mathbf{w}) \geq p_{\mathbf{y}|\mathbf{x}}(\mathbf{y}|\mathbf{x}) \middle| \mathbf{x},\mathbf{y} \right]$$
        for $\mathbf{w} \sim p_\mathbf{x}$ that is independent of $\mathbf{x}$. In terms of the information density $i(\mathbf{x};\mathbf{y}) \triangleq \ln{\left( \frac{p_{\mathbf{y}|\mathbf{x}}(\mathbf{y}|\mathbf{x})} {p_\mathbf{y}(\mathbf{y})} \right)}$, $g(\mathbf{x},\mathbf{y})$ may be also expressed as
            \begin{equation} \label{eq:g_xy} 
                g(\mathbf{x},\mathbf{y}) = P\left[i(\mathbf{w};\mathbf{y}) \geq i(\mathbf{x};\mathbf{y}) \middle| \mathbf{x},\mathbf{y} \right].
            \end{equation}
    \end{thm}

    \begin{proof}
        See \cite[Theorem 16]{polyanskiy2010channel}.
    \end{proof}

While the RCU bound works for arbitrary input distributions, we consider the average power constraint $\mathbb{E}\{|\tilde{x}_n|^2\}=1$ as it allows us to choose the capacity-achieving complex normal distribution for the input which greatly simplifies the bound. It should be noted that both the average power constraint and the constant power constraint yield the same PSD for the FTN signal. The RCU bound for the considered FTN signaling is given in the following theorem.

    \begin{thm}[RCU for FTN] \label{thm:FTN_RCU}
        Let $\mathcal{X}^2(k,\nu)$ denote the complex noncentral chi-square distribution. Let $V_n(\tilde{y}_n) \sim \mathcal{X}^2(1,|\tilde{y}_n|^2)$ for $\tilde{y}_n\in\mathbb{C}$ be independent in $n=0,1,\cdots,N-1$. For some target rate $R$ in bps/Hz, the RCU bound for the FTN channel model \eqref{eq:y_tilde} under the average input power constraint, $\mathbb{E}\{|\tilde{x}_n|^2\}=1, \forall n$, is given by
            \begin{equation} \label{eq:FTN_RCU_Pe}
            \boxed{ \begin{aligned}
        	P_e \leq \mathbb{E}_{\tilde{\mathbf{x}},\tilde{\mathbf{y}}}       [ &\min\{1,(2^{\Omega R}-1) \\
                &\cdot P\left[\sum_{n=0}^{N-1} \frac{1}{\sigma_n^2} V_n(\tilde{y}_n) \leq \mu(\tilde{\mathbf{x}},\tilde{\mathbf{y}}) \middle| \tilde{\mathbf{x}},\tilde{\mathbf{y}} \right] \Biggr\} \Biggr],
            \end{aligned} }
            \end{equation}
        
        \noindent where $\tilde{\mathbf{x}} \sim \mathcal{CN}(\mathbf{0},\mathbf{I})$ and $\tilde{\mathbf{y}}=\tilde{\mathbf{x}}+\tilde{\mathbf{z}}$ with $\tilde{\mathbf{z}} \sim \mathcal{CN}(\mathbf{0},\mathbf{D})$ having the covariance $\mathbf{D}=\text{diag}(\sigma_0^2,\cdots,\sigma_{N-1}^2)$, and $\mu(\tilde{\mathbf{x}},\tilde{\mathbf{y}}) \triangleq \sum_{n=0}^{N-1} \frac{1}{\sigma_n^2} |\tilde{x}_n-\tilde{y}_n|^2 = (\tilde{\mathbf{x}}-\tilde{\mathbf{y}})^\dagger \mathbf{D}^{-1} (\tilde{\mathbf{x}}-\tilde{\mathbf{y}})$.
    \end{thm}
    \begin{proof}
        See Appendix \ref{Appendix:ProofRCU}. 
    \end{proof}

As with the MC bound, computing the RCU bound is numerically challenging for moderate to large values of TBP and SNR (due to CDF in \eqref{eq:FTN_RCU_Pe} being small and often below the dynamic range of the computing precision). Similar to the MC bound, fortunately, we can accurately approximate the RCU bound (e.g., using methods that approximate CDF of sum of independent random variables – see Appendix \ref{Appendix:Approx}).\footnote{It is possible to apply an asymptotic expansion on the RCU bound, similar to the expansion applied to the MC bound in Appendix \ref{Appendix:asym_expansion_MC}. This yields 
    $R_\text{RCU} = C_\text{NA} - \sqrt{\frac{V_\text{RCU}}\Omega} \log_2(e) Q^{-1}(P_e) + \mathcal{O}\left(\frac{1}\Omega\right),$
where $V_\text{RCU} \triangleq \frac{1}{\Omega}\sum_{n=1}^{N} \frac{2}{1+\sigma_n^2}$. We omit the proof and instead refer the reader to the steps outlined in \cite{durisi2020lecture} using the Berry-Esseen theorem. Comparing this expansion to $R_{\text{NA}}$ from Proposition \ref{prop:FTN_NA}, we note that the second order term is larger due to $V_\text{RCU}=V_\text{NA}+\frac{1}{\Omega}\sum_{n=1}^{N} \frac{1}{(1+\sigma_n^2 )^2} > V_\text{NA}$ and the third order term, $\frac{\log_2(\Omega)}{2\Omega}$, is missing and instead contained in $\mathcal{O}(\frac{1}{\Omega})$. The larger second order term is a consequence of using \textit{i.i.d.} Gaussian symbols in deriving the RCU bound (which is a suboptimal choice in the FBL regime – see \cite{scarlett2016dispersion} for a similar observation). This implies that the derived RCU is in general less tight than the MC bound. Nevertheless, in our numerical results, the RCU bound is shown to provide tractable and reasonably tight lower-bound to MCCR in the finite TBP regime in most scenarios of interests.}

\section{Numerical Results} \label{section:NumResults}

\begin{figure}
    \centering
    \begin{subfigure}{\linewidth}
        \centering
        \includegraphics[width=\linewidth]{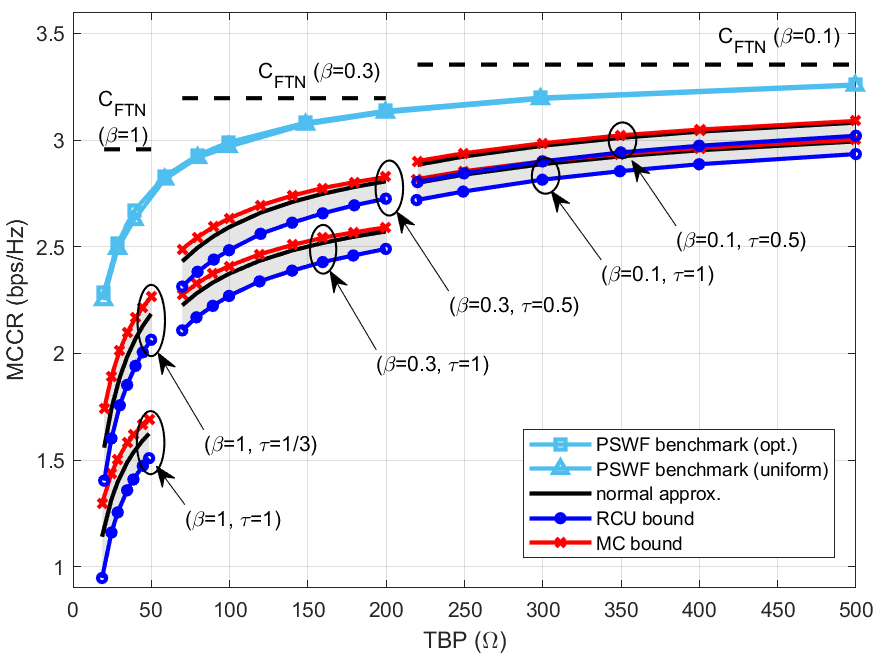}
        \caption{SNR $\rho=10\text{ dB}$}
    \end{subfigure}
    \begin{subfigure}{\linewidth}
        \centering
        \includegraphics[width=\linewidth]{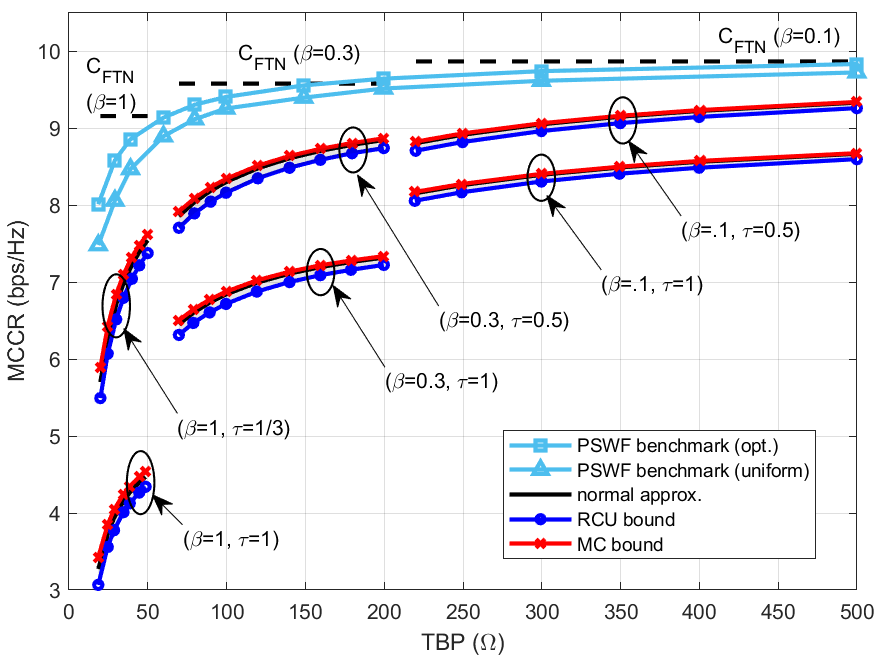}
        \caption{SNR $\rho=30\text{ dB}$}
    \end{subfigure}
    \caption{MCCR of FTN signaling in the finite TBP regime using RRC pulses with roll-off $\beta=1$ for $\Omega<50$, $\beta=0.3$ for $\Omega=(50,200$), and $\beta=0.1$ for $\Omega=(200,500)$.}
    \label{fig:MCCRvsTBP}
\end{figure}

Fig. \ref{fig:MCCRvsTBP} depicts MCCR of FTN signaling with the target BLER $P_e=10^{-3}$ and the OOB constraint $\epsilon_W=10^{-4}$. The plot includes NA, the MC upper-bound, and the RCU lower-bound at SNR $\rho=10$ dB and $30$ dB. For these FTN results, we assume RRC pulses that are time-truncated to meet the OOB constraint, having varying roll-off factors for different ranges of TBPs: i.e., $\beta=1$ for $\Omega\leq50$, $\beta=0.3$ between $50<\Omega\leq200$, and $\beta=0.1$ for $\Omega>200$ (these choices of $\beta$ lead to near-optimal MCCR for the respective TBP ranges as will be further elaborated in section \ref{subsection:pulsedesign}). The results are compared against the theoretically optimal PSWF benchmark, which is defined in Appendix \ref{Appendix:PSWF}. The MCCR ($R_{\text{NA}}$) of the PSWF benchmark with uniform or optimal symbol power allocations are plotted in Fig. \ref{fig:MCCRvsTBP}.

The figure shows that FTN signaling provides a significant MCCR improvement over the conventional Nyquist rate signaling (denoted by $\tau=1$) across all considered TBPs. Notably, these gains are more pronounced in the low-TBP regime and when SNR is higher. Furthermore, the performance of the FTN system closely approaches the theoretical PSWF benchmark in all TBPs. The results also confirm that the MC and RCU bounds are remarkably tight, and the NA consistently falls between them, validating NA as a simple yet accurate approximation to MCCR.

\begin{figure}
    \centering
    \includegraphics[width=\linewidth]{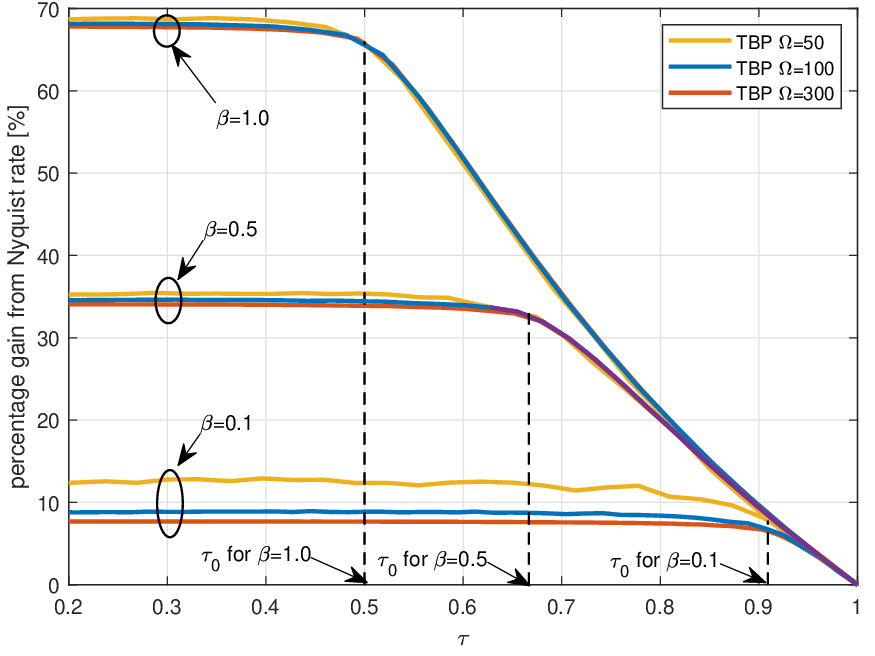}
    \caption{Percentage MCCR gain, $\frac{R_\text{NA}(\tau)-R_\text{NA}(1)}{R_\text{NA}(1)}\times100\%$, versus $\tau$ at various TBPs for truncated RRC pulses with roll-off $\beta$.}
    \label{fig:percentgain}
\end{figure}

To further quantify the relationship between the rate gain with the signaling rate, Fig. \ref{fig:percentgain} plots the percentage MCCR gain over the Nyquist baseline, defined as $(R_\text{NA}(\tau)-R_\text{NA}(1))/R_\text{NA}(1)$ as a function of $\tau$ at SNR $\rho=30 \text{ dB}$ and TBP $\Omega=\{50,100,300\}$ (other parameters remain unchanged). The plot shows that as $\tau$ is reduced, the MCCR gains increase steadily until they reach plateaus of approximately $10\%$, $35\%$ and $70\%$ for $\beta = 0.1, 0.5, \text{ and }1$, respectively. It is also interesting to observe that reaching these plateaus, particularly in the low-TBP regime, requires the time-acceleration factor to be strictly below the capacity-optimal limits, $\tau_0\triangleq1/(1+\beta)$ (illustrated as vertical lines in Fig. \ref{fig:percentgain}). The degree with which $\tau$ needs to be below $\tau_0$ to reach the plateaus, along with other FTN system design considerations, are investigated next. 

\section{FTN System Design} \label{section:FTNdesign}
This section discusses several design criteria for FTN systems when operating in the finite TBP regime. In particular, we identify $\tau$ needed to reach the maximum number of signaling dimensions, pulse selection based on TBP, and FTN coding design that can perform near the MCCR limit. We focus on the time-limited case with the OOB power constraint, but similar conclusions can be made for the bandlimited case. 

\subsection{Signaling rate} \label{subsection:PSWFbenchmarks}

The maximum number of signaling dimensions in a given TBP may be characterized by the number of PSWFs that can be transmitted subject to an OOB power constraint. This is formally defined as follows:

\begin{ndef} \label{ndef:maxDoF}
    The maximum number of signaling dimensions in a TBP $\Omega$ with an OOB power less than $\epsilon_W$ of the total power is the largest integer $N$ such that
        $$1 - \frac{1}{N} \sum_{n=0}^{N-1} \mu_{\Omega,n} \leq \epsilon_W,$$ 
    where $\mu_{\Omega,n}$ is the eigenvalue corresponding to the $n$-th PSWF, $\psi_{\Omega,n}(t)$, as defined in Appendix \ref{Appendix:PSWF}. This eigenvalue also represents the energy concentration of the corresponding normalized and truncated PSWF, $\phi_{\Omega,n}(t)$, within the frequency band $|f|<W$. 
\end{ndef}

The maximum number of signaling dimensions, denoted $N^*$, is a function of both $\Omega$ and $\epsilon_W$. For convenience, we express this as $N^*=\Omega-\eta$, where $\eta$ represents the dimensional loss relative to TBP. It is well-known that $N^*/\Omega \to 1$ as $\Omega\to\infty$, a result known as the \textit{2WT theorem} \cite{slepian1983some}. In the finite TBP regime, $\eta$ is typically small for practical values of $\epsilon_W$ (e.g., for $\epsilon_W=10^{-4}$ and $\Omega < 500$, $3<\eta<5$). As a rule of thumb, $\eta\approx-\log_{10}(\epsilon_W)$ for small $\Omega$, and this value tends to decrease slowly as $\Omega$ increases.  

One of the key benefits of FTN signaling is its ability to increase the number of parallel channels for a given TBP. To utilize the maximum number of signaling dimensions, the time-acceleration factor $\tau$ can be set to yield $N^*$ channels as follows:
    \begin{equation} \label{eq:taustar}
        \boxed{ \tau^*\triangleq\frac{T_x-T_p}{(N^*-1)T}=\frac{\Omega-c}{\Omega-\eta-1} \tau_0,}
    \end{equation}
due to \eqref{eq:N} and substituting $T=\frac{1+\beta}{2W}$, $c \triangleq 2WT_p$, and $\tau_0 \triangleq \frac{1}{1+\beta}$.  

\begin{figure}
    \centering
    \includegraphics[width=\linewidth]{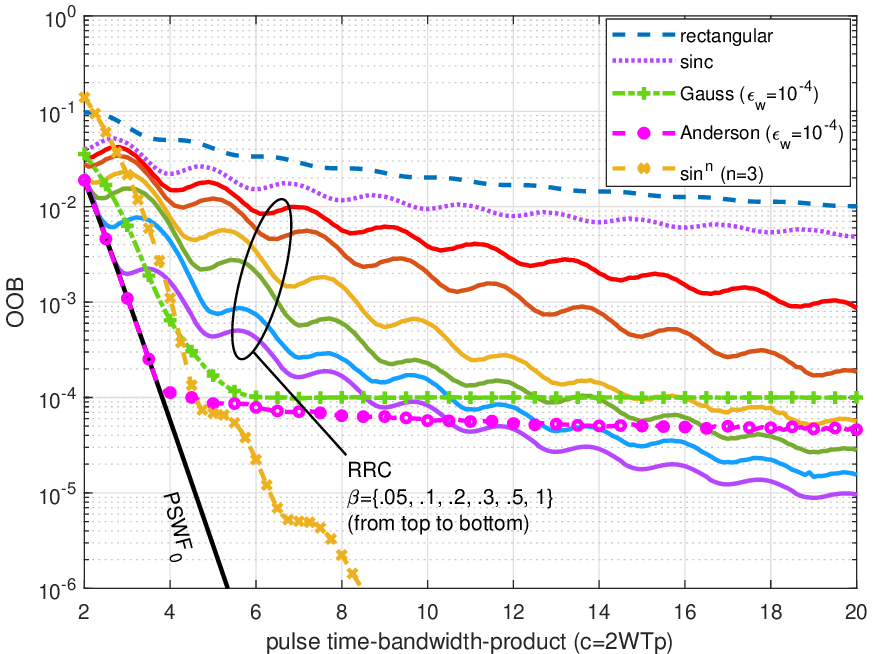}
    \caption{OOB energy of various pulse shapes in terms of $c=2WT_p$}
    \label{fig:OOBE_vs_c}
\end{figure}

We see from \eqref{eq:taustar} that $\tau^*$ approaches $\tau_0$ in the asymptotic TBP regime (i.e., as $\Omega \to \infty$). In the finite TBP regime, however, $\eta$ is typically smaller than $c$ for practical ranges of $\epsilon_W$, which implies $\tau^*<\tau_0$ is required to utilize all $N^*$ dimensions. The minimum $c$ needed to meet a given OOB power constraint can be inferred from Fig. \ref{fig:OOBE_vs_c}, which plots the OOB energy versus $c$ for various pulses. For instance, $c \geq 10.52$ for the RRC pulse with $\beta=0.5$ when $\epsilon_W=10^{-4}$.

\begin{figure}
    \centering
    \includegraphics[width=\linewidth]{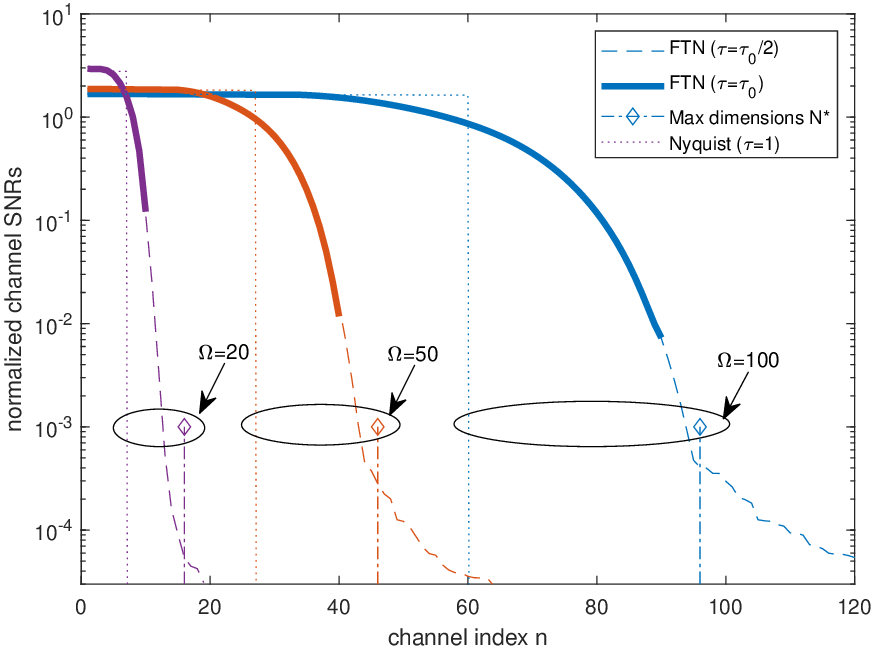}
    \caption{Normalized SNRs of the parallel channels ($\sigma_n^{-2}/\rho$) with truncated RRC pulses with $\beta=.5$ and $\epsilon_W=10^{-4}$.}
    \label{fig:channelSNRs}
\end{figure}

Fig. \ref{fig:channelSNRs} plots the normalized SNRs of the parallel channels ($\sigma_n^{-2}/\rho$) for FTN systems with various $\tau$ values. As shown in solid lines, using $\tau_0$ activates only a partial number of channels, especially in low-TBP regimes, indicating that there is a benefit of selecting $\tau < \tau_0$ in the finite TBP regime. Setting $\tau=\tau^*$ yields $N^*$ channels as also indicated in the figure. The levels of these additional channels scale linearly with $\rho$ and thus the rate gains are higher in the high SNR. There is a diminishing return on rate gains for $\tau<\tau^*$, as the additional channels beyond $N^*$ possess diminishing SNRs and thus contribute little to the overall rate. 

\subsection{Time-limited pulse design} \label{subsection:pulsedesign}

An optimal time-limited pulse must satisfy several criteria: it should generate minimal ISI at the Nyquist rate, adhere to a given OOB power constraint, and maximize the MCCR when used as a base pulse of FTN system.

Various pulse designs have been proposed for FTN signaling, ranging from classical waveforms like the RRC, Gaussian, and PSWFs to pulses optimized for specific performance metrics. Liveris and Georghiades \cite{liveris2003exploiting} established the existence of the Mazo limit for RRC pulses, demonstrating that they can be packed beyond the Nyquist rate without reducing the minimum Euclidean distance. Le et al. \cite{le2014practical} evaluated several $T$-orthogonal pulses including RRC, Phydyas, and extended Gaussian functions regarding their bit error rate, peak-to-average power ratio, and spectral efficiency. Rusek et al. \cite{rusek2008faster} explored short, finite pulses such as the half-cycle sinusoid and Halpern's finite Nyquist pulse. However, the lack of a common OOB energy baseline makes direct comparisons difficult. The use of PSWFs for FTN appears in the early work of Foschini \cite{foschini1984contrasting} and Anderson and Rusek \cite{anderson2007optimal}. Notably, PSWFs and Gaussian pulses were identified to provide superior performance to RRC pulses when packet length is short \cite{anderson2007optimal}. 

More recent literature focuses on optimized pulse design under some performance metrics. Anderson \cite[Ch. 7]{anderson2017bandwidth} explored several optimized pulses, including a modified principal PSWF, that simultaneously satisfy OOB and OOI requirements; however, these idealized pulses have infinite time and frequency support. Jaffal and Alvarado \cite{jaffal2022pulses} proposed a weighted combination of normalized, time-truncated PSWFs, where weights are optimized to minimize residual ISI power (specifically power beyond the equalizer's memory) under OOB constraints. Similarly, Jaffal and Abou-Faycal \cite{jaffal2019achievable} utilized linear combination of PSWFs to maximize capacity in bandlimited channels. Diverging from orthogonal designs, Milojkovi\'c et al. \cite{milojkovic2022pulseshaping} introduced non-orthogonal pulses formed by linear combination of RRC pulses, optimized to maximize an information rate upper bound. Additionally, Makarov et al. \cite{makarov2020optimizing} proposed time-limited pulses based on Fourier-series (FS) expansion, optimized for minimum OOB power subject to specific autocorrelation constraints. Despite these advancements, the time-truncated RRC pulse remains the common choice in the FTN literature due to its versatility and wide adoption in communication standards.

In this section, we develop an optimized pulse design based on FS expansion, similar to the approach in \cite{makarov2020optimizing}. Consider an $m$-th order FS approximation of a real-valued, even-symmetric pulse $p(t)$:
    $$p(t)=c_0+\sum_{k=1}^{m-1} 2|c_k|\cos{\left(k\frac{2\pi}{T_p}t\right)} \text{ for } |t|\leq \tfrac{T_p}{2},$$
where $\{c_k\}$ are the FS coefficients. The corresponding Fourier transform is
    \begin{equation*} \begin{split}
        \hat{p}(f)& = T_p\bigg[c_0 \text{sinc}(fT_p) \\
        &+ \sum_{k=1}^{m-1} |c_k|\left( \text{sinc}\!\left((f-\tfrac{k}{T_p})T_p \right) + \text{sinc}\!\left((f+\tfrac{k}{T_p})T_p \right) \right) \bigg],
    \end{split} \end{equation*}
and the autocorrelation function is 
    \begin{equation*} \begin{split}
        h_n 
         &= c_0^2(T_p-n\tau T) - \sum_{k=1}^{m-1} 4c_0|c_k|\frac{(-1)^k}{k\omega_p}\sin(kn\omega_p\tau T) \\
        &+\sum_{k=1}^{m-1}\sum_{r=1}^{m-1} 2|c_k||c_r|g_{k,r}, \text{ for } |n|\leq \left\lfloor {\tfrac{T_p}{\tau T}} \right\rfloor,
    \end{split} \end{equation*}
where $\omega_p\triangleq \frac{2\pi}{T_p}$ and $g_{k,r}$ is defined as
    \begin{equation*} \begin{split}
        \begin{cases}
            \frac{2(-1)^{k-r}}{(k^2-r^2)\omega_p}(r\sin(r\omega_p n\tau T)-k\sin(r\omega_p n\tau T)), \text{ if } k \neq r \\
            \cos(k\omega_p n\tau T)(T_p\!-\!n\tau T)-\frac{1}{k\omega_p}\sin(k\omega_p n\tau T), \text{ if } k=r.
        \end{cases}
    \end{split} \end{equation*}

A unique feature of the FS-based design is that the coefficients $\{c_k\}$ directly shape the pulse spectrum at discrete frequencies $f_k=\frac{k}{T_p}$, since $\hat{p}(f_k)=|c_k|T_p$. Therefore, optimizing the coefficients may be interpreted as shaping the spectrum at these points, with the response between points interpolated by sinc functions. To satisfy the OOB power requirement, the coefficients $\{c_k\}$ for $|k|>\left\lceil WT_p \right\rceil$, which correspond to frequencies beyond the band limit $W$ Hz, must be driven close to zero. It should be noted that the resulting pulse is not ISI-free, and must to be designed subject to an autocorrelation requirement, e.q., $\max_n|h_n|<K_0$ at $\tau=1$. 

Following \cite{makarov2020optimizing}, we optimize the FS coefficients, but instead of minimizing OOB power, we maximize the first-order NA of MCCR, $C_\text{NA}$, at the Nyquist rate with $\tau=1$, subject to both OOB power and autocorrelation constraints. The optimization is performed for every TBP and SNR pair. The problem is formally stated as follows: 

Given $\{\Omega, \epsilon_W, \rho, K_0, W\}$, for $T_p=\{\frac{1}{2W},\frac{1.5}{2W},\cdots,\frac{20}{2W}\}$ or until $T_p=T_x$, $m=\lceil WT_p \rceil+2$, and $\tau=1$, we solve
\begin{equation*} \begin{aligned}
    \max_{\substack{c_0,|c_1|,\cdots,|c_{m-1}|,\\\frac{1}{2W}\leq T<T_p}} &\quad \frac{1}{\Omega}\sum_{m=0}^{N-1}{\log_2(1+\rho\lambda_n)}, \\
    \text{s.t.} \quad & T_p\left(c_0^2+\sum_{k=1}^{m-1}2|c_k|^2\right)=1 \text{ (unit energy)}\\
    & 1-2\int_0^W{|\hat{p}(f)|^2df} \leq \epsilon_W \text{ (OOB)}\\
    &\max_{n=\{1,\cdots,L\}}{|h_n|}<K_0,\, L=\left\lfloor \tfrac{T_p}{T} \right\rfloor \text{ (autocorr.)},
\end{aligned} \end{equation*}
where $\lambda_n$ is the $n$-th eigenvalue of the FTN matrix $\mathbf{H}$ when $\tau=1$ (which is non-diagonal because $p(t)$ is not ISI-free), and $N=\left\lfloor (T_x-T_p)/T+1 \right\rfloor$.

The solution to this problem is a unit-energy, time-limited pulse with an OOB power below $\epsilon_W$ of the total power and a maximum ISI below $K_0$ when signaled at the Nyquist rate $1/T$. We used the MATLAB optimization solver \textit{fmincon} to search for the solutions. Table \ref{tab:optpulFScoeff} lists three resulting pulses for $c=2WT_p=\{4,6,10\}$ along with their corresponding FS coefficients $\{c_k\}$ and the Nyquist symbol times $T$. These pulses are designed for the ranges of $\Omega$ shown in the table. 

\begin{figure} \centering
    \includegraphics[width=1\linewidth]{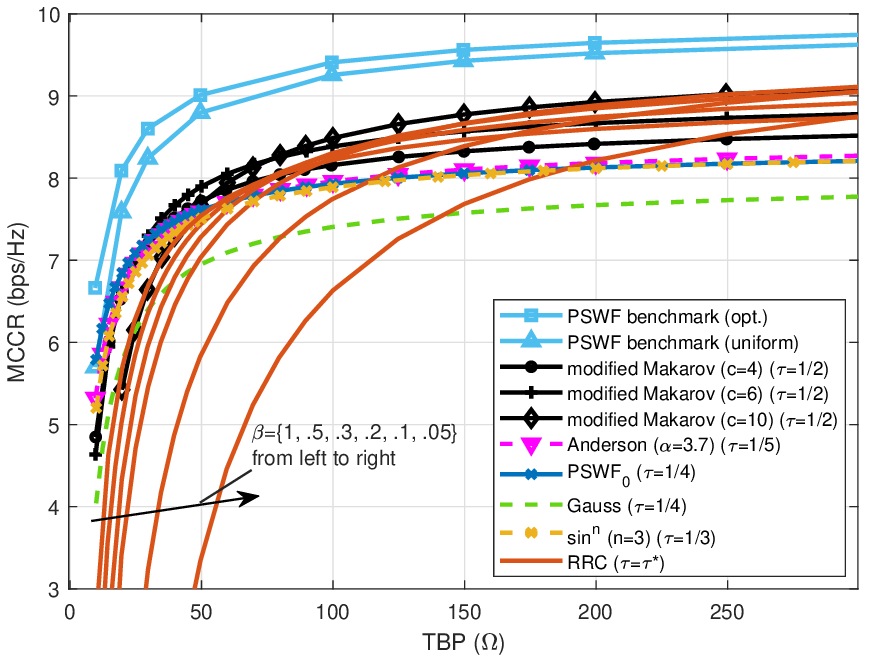}
    \caption{MCCRs ($R_{\text{NA}}$) of FTN signaling with various time-limited pulse shapes with $\epsilon_W=10^{-4}$ at $\rho=30 \text{ dB}$. PSWF benchmarks are also shown.}
    \label{fig:pulseMCCRvsc}
\end{figure}

\begin{table}
    \centering
    \begin{tabular}{c| c| c| c | c}\toprule
          $c$  &   $T$         &      $c_0$      &        $|c_k|,\ k\geq1$        & $\Omega$ best used    \\  \midrule
          4  &  2.2692  &  0.4106  & \{0.2017, 0.0063, 0.0002\} & $\leq 10$ \\ \hline
          6  &  1.6923  &  0.1931  & \makecell{\{0.2202, 0.1271, 0.0079, \\0.0004\}} & $(10,50)$ \\ \hline
          10  &  1.2154  &  0.1068  & \makecell{\{0.1136, 0.0934, 0.1245, \\0.0846, 0.0047, 0.0005\}} & $(50,200)$\\ \bottomrule
    \end{tabular}
    \caption{Optimized FS coefficients for modified Makarov \cite{makarov2020optimizing}; designed with parameters $\epsilon_W=10^{-4}, \rho=20 \text{ dB}, K_0=0.1, W=0.5$ (different $W$ may be obtained by time scaling)}    
    \label{tab:optpulFScoeff}
\end{table}

We now compare the performance of the optimized pulse against other common pulses. Fig. \ref{fig:pulseMCCRvsc} plots $R_\text{NA}$ for FTN systems using various pulses, alongside the PSWF benchmarks as detailed in Appendix \ref{Appendix:PSWF}. The considered pulses include our optimized pulse from Table \ref{tab:optpulFScoeff} (labeled "modified Makarov"), time-truncated RRC pulses with various $\beta$, Anderson's modified PSWF \cite{anderson2017bandwidth}, the principal PSWF pulse $\phi_{c,0}(t)$ (labeled "PSWF$_0$"), Gaussian pulse, and half-cycle sinusoid raised to $n$-th power (labeled "$\sin^n$").\footnote{Anderson's pulse is given by $p(t)=\alpha_1 \psi_{c,0}(t) \text{ for } |t|\leq T/2$ and $p(t)=\alpha_2 \psi_{c,0}(t) \text{ for } |t|>T/2$, where $\alpha_1$ and $\alpha_2$ are chosen so that OOB and OOI are both equal to $\epsilon$. See \cite{anderson2017bandwidth} for details. Anderson's pulse is further time-truncated to $|t|<T_p/2$ and energy normalized in our work. Gauss is a time-truncated and energy normalized Gaussian function with the parameter $\sigma=Q^{-1}\left(\epsilon_W/2\right)/(2\sqrt{2}\pi W)$, chosen to make OOB equal to $\epsilon_W$. Half-cycle sinusoid ($\sin^n$) is given by $\cos^n(\pi t/T_p)$ for $|t|\leq T_p/2$ and $n\in \mathbb{Z^+}$, which is further normalized to have unit energy.} For each pulse, the minimum pulse-width (or the minimum $c=2WT_p$) needed to meet the OOB constraint $\epsilon_W=10^{-4}$ is obtained from a plot of OOB energy versus $c$ in Fig. \ref{fig:OOBE_vs_c}.%
\footnote{We use the conventional definition of $T=\tfrac{1+\beta}{2W}$ for RRC. The non-Nyquist pulses \{Anderson, $\text{PSWF}_0$, Gauss, and $\text{sin}^n$\}, however, do not have a well-defined Nyquist symbol time. For these pulses, we use the generalized definition of Nyquist symbol time proposed in \cite{zhou2012generalized}. Specifically, the normalized bandwidth is first defined as $2B_\text{norm}=\tfrac{\int_{-W}^{W} |\hat{p}(f)|^2df}{\max_f|\hat{p}(f)|^2}$, which is the base-width of a rectangle whose height equals the peak of the pulse spectrum and whose area equals the in-band energy. The generalized Nyquist symbol time is then defined as $T=\tfrac{1}{2B_\text{norm}}$. The roll-off factor is also generalized as $\beta=\tfrac{W-B_\text{norm}}{B_\text{norm}}$. These definitions reduce to the conventional definitions when applied to RRC and sinc pulses.}

Among all considered pulses, the principal PSWF yields the highest MCCR at very low TBP $(\Omega<10)$. However, it is quickly outperformed by other pulses, making it a poor choice for a general FTN base pulse except in a very low-TBP regime or under extremely strict OOB constraints. For the RRC family, a roll-off of $\beta=1$ is the best at low TBPs, whereas smaller values of $\beta$ are superior at higher TBPs, with the sinc pulse ($\beta\to0$) being asymptotically optimal. Our modified Makarov pulse outperforms the RRC pulses for low and moderate TBPs ($\Omega<200$) and performs closer to the PSWF benchmarks. The remaining pulses exhibited lower performance than either the principal PSWF or the RRC pulses across for all TBPs considered. 

In summary, the principal PSWF and the sinc pulse are optimal at the two extremes of the TBP range. In the finite TBP regime, the RRC pulses with $\beta$ selected according to TBP effectively bridge the performance gap, making them a strong candidate for FTN systems. To achieve performance closer to the theoretical PSWF benchmarks, optimized pulse designs, such as the modified Makarov pulse proposed here, can be employed. 

\subsection{Coding design} \label{subsection:coding}
A natural question following the MCCR analysis is whether practical FTN coding systems can reach the performance limits. In this section, we describe an FTN coding system that can perform within $1.3 \text{ dB}$ of the minimum BLER \eqref{eq:P_e} in a TBP as low as $\Omega\approx100$.

\begin{figure} \centering
    \includegraphics[width=1\linewidth]{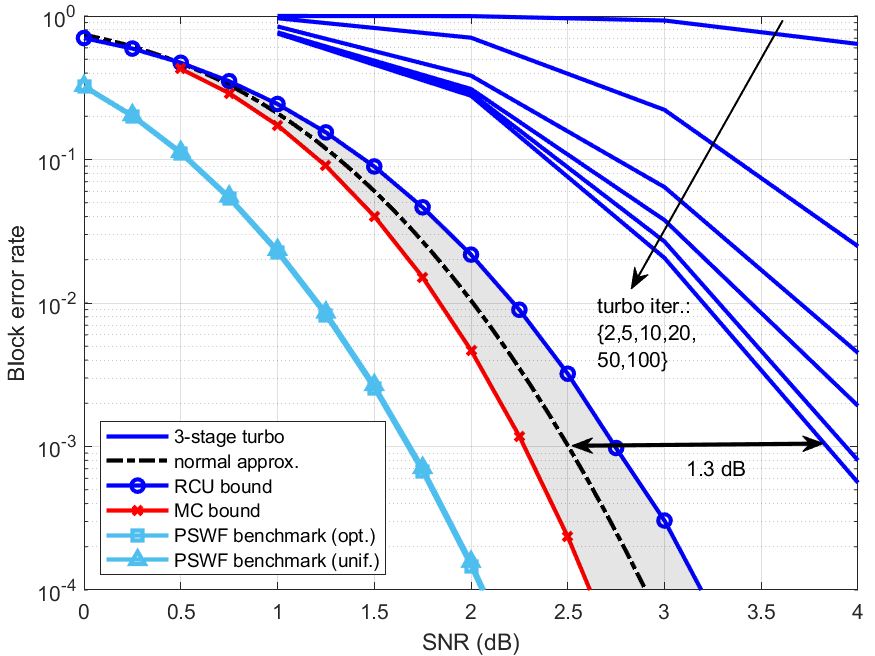}
    \caption{Performance of the three-stage FTN coding system along with the minimum BLER at TBP $\Omega=132$. Other parameters include $L=3$, $I_\text{in}=5$, and $I_\text{out}$ up to $100$.}
    \label{fig:BLER_3stageturbo}
\end{figure}

The FTN coding system is based on the three-stage turbo equalization \cite{sugiura2015frequency}, with small modifications to allow near-MCCR performance in the finite TBP regime. In the FTN transmitter, information bits are first encoded using a rate-$1/2$ recursive systematic convolutional (RSC) encoder, bit-interleaved, and then encoded using unity-rate recursive convolutional (URC) encoder. The same RSC and URC encoders with unit memory, as in \cite{sugiura2015frequency}, are used. The coded bits are bit-interleaved again and mapped to modulation symbols $\{x_n\}$ (e.g., QPSK), before being modulated into the FTN signal \eqref{eq:x(t)} and transmitted over the channel. The URC encoder and the FTN modulation together effectively creates a recursive ‘inner-code’ of a turbo code, which facilitates a turbo decoding loop with the RSC encoder as the ‘outer-code’. 

The FTN receiver implements three-stage turbo equalization between the FTN channel equalizer, URC decoder, and RSC decoder that iteratively exchange the extrinsic information of the coded bits. Instead of the low-complexity frequency-domain FTN equalizer considered in \cite{sugiura2015frequency}, we use the optimal maximum a posteriori (MAP) equalizer with $L$-memory that is operating on the trellis description of the FTN channel model \eqref{eq:y}. Leveraging on the fact that the FTN channel model is an instance of the Ungerboeck observation model, the MAP equalizer is implemented using the BCJR algorithm with modified branch metrics (see eq. (11) in \cite{rusek2015ungerboeck})\footnote{The computational complexity of the optimal MAP equalizer is $\mathcal{O}(M^LN)$ for $L$-memory equalizer with $M$ modulation order and $N$ symbols. The exponential complexity can be replaced by polynomial (or even linear) complexity using reduced-complexity FTN equalizers at some performance loss \cite{sugiura2015frequency}, \cite{bedeer2017very}. Fortunately, $L$ generally need not be large, since RRC pulses with $\beta$ close to 1 that is shown to be near-optimal in low TBP has fast temporal roll off and generates short ISI even when $\tau \ll 1$. As TBP increases, larger $\beta$ is needed but at the same time $\tau$ need not be small (both $\tau^*$ and $\tau_0$ get closer to 1) thus ISI remains short. Channel shortening technique \cite{fan2018mlse, li2020code} can be used to further lower $L$ if needed. Numerical simulations confirmed that $L=3$ was sufficient to perform near MCCR in most cases of interest.}.  The MAP equalizer and the URC decoder form an inner turbo loop and exchange the extrinsic information about the URC encoded bits. The URC decoder generates the extrinsic information of both the URC encoded bits and the RSC encoded bits. After the inner loop iterates for $I_\text{in}$ times, the outer turbo loop between the URC decoder and the RSC decoder is run, in which the extrinsic information about the RSC encoded bits is exchanged. The inner loop then restarts with the updated extrinsic information from the RSC decoder becoming the a prior information about the URC encoded bits. The outer loop iterates for $I_\text{out}$, with which the total number of turbo iteration is $I_\text{in}I_\text{out}$. See \cite{sugiura2015frequency} for more details.

It is known that turbo codes occasionally generate low-weight codewords, leading to an error floor in the bit-error rate curve \cite{takeshita2002frame}. To lower the error floor and to improve the BLER performance in the finite TBP regime, we use the $S$-random interleavers \cite{dolinar1995weight} with at least $S=\lfloor \sqrt{l/2} \rfloor$ separations between the bits in the interleaved sequence, where $l$ is the interleaver depth. The $S$-random interleavers are found numerically, and two different interleaving patterns are used (with the same $S$) in our system and are fixed throughout the simulation. 

Fig. \ref{fig:BLER_3stageturbo} shows BLER performance of the three-stage FTN coding system. We consider transmission of $128$ information bits. With the overall coding rate of 1/2 (note that no extra overhead is needed in the given construction) and QPSK modulation with Gray mapping, the packet size is $N=128$ symbols. We use truncated RRC with roll-off $\beta=1$ with OOB constraint $\epsilon_W=10^{-4}$ (which yields $c=8.57$ from Fig. \ref{fig:OOBE_vs_c}). The minimum TBP required to transmit this packet is given by $\Omega=132$, obtained from $\Omega=N^*+\eta$ with $N=N^*$ and $\eta\approx4$ for $\Omega<200$. We set the FTN factor $\tau=\tau^*=0.4859$ to pack $N^*=128$ symbols in the given TBP (in contrast, Nyquist rate signaling can only include $62$ symbols within this TBP). The resulting spectral efficiency of the FTN system is $N/\Omega=128/132\approx0.970$ bps/Hz. In the figure, the minimum BLER \eqref{eq:P_e} along with the corresponding MC and RCU bounds at $\Omega=132$ when the target rate is set as the spectral efficiency of the FTN system are also displayed. As seen in the figure, the three-stage FTN coding system can perform within $1.3 \text{ dB}$ from the minimum BLER at the target $P_e=10^{-3}$, demonstrating that the derived MCCRs can be closely achieved with practical FTN coding systems.

\section{Conclusion}

This work established a rigorous framework for analyzing FTN signaling in the finite TBP regime, clarifying its benefits for short packet communications. By deriving tight bounds on the MCCR, we demonstrated that FTN achieves significant rate gains over Nyquist signaling, with these advantages being most prominent at low TBP and high SNR. Alternatively, FTN can be leveraged to enhance reliability by lowering the BLER for a fixed coding rate.

From a design perspective, we identified the optimal time-acceleration factor needed to maximize the available signaling dimensions and showed that with optimized pulse shaping, FTN's performance can closely approach the theoretical benchmark set by PSWFs. Furthermore, we validated these theoretical gains with a practical three-stage turbo-equalization coding scheme that performs close to the MCCR bound, even for a TBP as low as $\Omega=132$. Further performance gain is expected by using more advanced and recent FTN receiver designs (e.g., \cite{li2020code}, \cite{yang20256g}) by increasing the computational complexity.

Collectively, these results confirm that FTN effectively reduces the performance penalty from finite-length coding by creating more signaling dimensions within a constrained TBP. The fundamental benefits identified are expected to extend to more complex environments, such as fading, MIMO, and multi-user channels, positioning FTN as a valuable technique for future communication systems \cite{li2026faster}.

\appendices

\section{Proof of Theorem \ref{thm:MC_FTN}} \label{Appendix:ProofMC}
\begin{proof}
    For the $N$-parallel Gaussian channel model, we have
        $$p_{\tilde{y}|\tilde{x}}(\tilde{\mathbf{y}}|\tilde{\mathbf{x}}) =
        \prod_{n=0}^{N-1} p_{\tilde{y}|\tilde{x}}(\tilde{y}_n|\tilde{x}_n) = \prod_{n=0}^{N-1} \frac{1}{\pi\sigma_n^2} e^{-\frac{1}{\sigma_n^2} |\tilde{y}_n - \tilde{x}_n|^2}.$$
    A judicious choice for $q_{\tilde{y}}(\tilde{\mathbf{y}})$ is the capacity-achieving output distribution: $\tilde{\mathbf{y}} \sim \mathcal{CN}(\mathbf{0},\mathbf{I}+\mathbf{D})$, where $\mathbf{D}$ is a diagonal covariance matrix of $\tilde{\mathbf{z}}$ with $n$-th diagonal entry equal to $\sigma_n^2$. Thus,
        $$q_{\tilde{y}}(\tilde{\mathbf{y}}) = \prod_{n=0}^{N-1} q_{\tilde{y}}(\tilde{y}_n) = \prod_{n=0}^{N-1} \frac{1}{\pi(1+\sigma_n^2)} \exp{\left(-\tfrac{1}{1+\sigma_n^2} |\tilde{y}_n|^2\right)}.$$ 
    Let \textit{mismatched} information density be defined as       $$i_{\text{M}}(\tilde{x}_n;\tilde{y}_n) \triangleq \ln\left( \frac{p_{\tilde{y}|\tilde{x}}(\tilde{y}_n | \tilde{x}_n)} {q_{\tilde{y}}(\tilde{y}_n)}\right).$$ 
    It can be simplified as
        \begin{equation} \label{eq:mism_infdensity}
            i_{\text{M}}(\tilde{x}_n;\tilde{y}_n) = \ln\left(1+\frac{1}{\sigma_n^2}\right) - 
            \frac{|\tilde{y}_n - (1+\sigma_n^2)\tilde{x}_n|^2}{\sigma_n^2 (1+\sigma_n^2)} + |\tilde{x}_n|^2.
        \end{equation}
    The FA probability is then
        \begin{equation*} \begin{split}
            P_{\text{FA}}(\tilde{\mathbf{x}}&,\lambda) 
            = P\Biggl[\frac{1}{N} \sum_{n=0}^{N-1} \ln\left(1+\frac{1}{\sigma_n^2} \right) \\ 
            & - \frac{1}{N} \sum_{n=0}^{N-1} \frac{|\tilde{y}_n - (1+\sigma_n^2)\tilde{x}_n|^2}{\sigma_n^2 (1+\sigma_n^2)} + 1 > \lambda \Bigg| \text{H}_0,\tilde{\mathbf{x}}\Biggr],
        \end{split} \end{equation*}
    
    \noindent due to $|\tilde{x}_n|^2=1$. Under the hypothesis $\text{H}_0$, $\tilde{y}_n \sim q_{\tilde{y}}(\tilde{y}_n) = \mathcal{CN}(0, 1+\sigma_n^2)$ and 
        \begin{equation} \label{eq:Un}
    	\left|\frac{\tilde{y}_n - (1+\sigma_n^2)\tilde{x}_n} {\sqrt{1+\sigma_n^2}}\right|^2 \sim \mathcal{X}^2(1,1+\sigma_n^2),
        \end{equation}
    where the noncentrality parameter, $1+\sigma_n^2$, is obtained by the assumption $|\tilde{x}_n|^2=1$. Defining $U_n$ as \eqref{eq:Un}, the FA probability, $P_{\text{FA}}(\tilde{\mathbf{x}},\lambda)$, may be simplified as
        $$ P\left[\frac{1}{N} \sum_{n=0}^{N-1} \frac{1}{\sigma_n^2} U_n < - \lambda + 1 + \frac{1}{N} \sum_{n=0}^{N-1} \ln\left(1+\frac{1}{\sigma_n^2}\right) \right].$$
    The above is independent of $\tilde{\mathbf{x}}$, hence $P_{\text{FA}} (\tilde{\mathbf{x}},\lambda) = P_{\text{FA}}(\lambda)$. 
    
    As for the MD probability, note that under the hypothesis $\text{H}_1$, $\tilde{y}_n \sim p_{\tilde{y} | \tilde{x}} (\tilde{y}_n | \tilde{x}_n) = \mathcal{CN}(\tilde{x}_n, \sigma_n^2)$ and
        \begin{equation} \label{eq:Vn}
            \left|\frac{\tilde{y}_n - (1+\sigma_n^2)\tilde{x}_n} {\sqrt{\sigma_n^2}}\right|^2 \sim \mathcal{X}^2(1,\sigma_n^2),
        \end{equation}

    \noindent where the noncentrality parameter is again obtained by the assumption $|\tilde{x}_n|^2=1$. Defining $V_n$ as \eqref{eq:Vn}, the MD probability, $P_{\text{MD}}(\tilde{\mathbf{x}},\lambda)$, may be simplified as
        $$ P\left[\frac{1}{N} \sum_{n=0}^{N-1} \frac{1}{1+\sigma_n^2} V_n > -\lambda + 1 + \frac{1}{N} \sum_{n=0}^{N-1} \ln{\left(1+\frac{1}{\sigma_n^2}\right)} \right].$$
    Again, the MD probability is independent of $\tilde{\mathbf{x}}$, hence $P_{\text{MD}}(\tilde{\mathbf{x}},\lambda)=P_{\text{MD}}(\lambda)$. Finally, without loss of generality, we may redefine $\lambda$ as the terms appearing on RHS of the inequalities in both $P_\text{FA}$ and $P_\text{MD}$. Substituting these into Theorem \ref{thm:PPV_MC} and changing the units yields the desired result.
\end{proof}

\section {Asymptotic expansion of the MC bound and proof of Proposition \ref{prop:FTN_NA}} \label{Appendix:asym_expansion_MC}

Under the settings of Theorem \ref{thm:MC_FTN}, we first relax the MC bound using the following inequality on the FA probability \cite[eq. (106)]{polyanskiy2010channel}:
    $$P_{\text{FA}}(\lambda) \leq \sup_{\gamma>0} \left\{e^{-N\gamma} \left( (1-P_e)-(1-P_{\text{MD}}(\gamma)) \right) \right\}.$$
This bound assumes that the FA and MD probabilities are both independent of the transmitted codeword. Applying the above bound to Theorem \ref{thm:PPV_MC} results in a generalized version of Verdú-Han converse bound (in bpcu) \cite{verdu1994general}, \cite{durisi2020lecture}:
    \begin{equation} \label{eq:VH_bound}
        R \leq \inf_{\gamma>0} \left\{ \gamma \log_2(e) - \frac{1}{N} \log_2\left(P_{\text{MD}}(\gamma) - P_e\right) \right\}.
    \end{equation}

\noindent For the FTN channel, recall that the MD probability is
    $$P_{\text{MD}}(\gamma) = P\left[ \sum_{n=0}^{N-1} i_{\text{M}}(\tilde{x}_n; \tilde{y}_n) \leq N\gamma \middle| \text{H}_1,\tilde{\mathbf{x}} \right],$$

\noindent where $i_{\text{M}}(\tilde{x}_n; \tilde{y}_n)$ is the \textit{mismatched} information density and given by \eqref{eq:mism_infdensity} for the FTN channel. We will need its first three moments, which are given in the lemma below. 

    \begin{lem} \label{lem:moments}
        (moments of mismatched information density): Let $|\tilde{x}_n|=1$. The mean and the variance of the mismatched information density $i_{\text{M}}(\tilde{x}_n; \tilde{y}_n)$ under $\tilde{y}_n \sim p_{\tilde{y} \mid \tilde{x}} = \mathcal{CN}(\tilde{x}_n, \sigma_n^2)$ and for a fixed $\tilde{x}_n$ are
            \begin{equation*} 
            \begin{split}
                &c_n \triangleq \mathbb{E}_{\tilde{y}_n} [i_{\text{M}}(\tilde{x}_n; \tilde{y}_n)] = \ln\left(1+\tfrac{1}{\sigma_n^2}\right), \\ &v_n  \triangleq \text{var}_{\tilde{y}_n} [i_{\text{M}}(\tilde{x}_n; \tilde{y}_n)] = 1 - \frac{1}{(1+\frac{1}{\sigma_n^2})^2}.
            \end{split}
            \end{equation*}
        The third absolute central moment, $\theta_n \triangleq \mathbb{E}_{\tilde{y}_n} [|i_{\text{M}}(\tilde{x}_n; \tilde{y}_n) - c_n|^3 ]$, can be bounded as
		$$v_n^{3/2} \leq \theta_n \leq \sqrt{27} v_n^{3/2}.$$
    \end{lem}

    \begin{proof} 
        Both $c_n$ and $v_n$ are straightforward. $\theta_n$ may be bound using the power-norm inequality:
            $$(\mathbb{E}_{\tilde{y}_n} [|i_{\text{M}}(\tilde{x}_n;\tilde{y}_n)-c_n|^2 ])^{\frac{1}{2}} \leq \theta_n^{\frac{1}{3}} \leq (\mathbb{E}_{\tilde{y}_n} [|i_{\text{M}}(\tilde{x}_n;\tilde{y}_n)-c_n|^4])^{\frac{1}{4}}.$$
        The proof is then complete by noting that the fourth central moment may be upper-bounded by $9v_n^2$.
    \end{proof}

    \begin{cor}
        Let $B \triangleq {(\frac{1}{\Omega} \sum_n \theta_n)} / {(\frac{1}{\Omega} \sum_n v_n )^{3/2}}$ . As TBP $\Omega$ grows, $B$ behaves as $\mathcal{O}(1)$ in the FTN channel.
    \end{cor}

We are now ready to present a proof of Proposition \ref{prop:FTN_NA}.

\begin{proof}[Proof of Proposition \ref{prop:FTN_NA}]
We apply the Berry-Esseen central limit theorem (CLT) to lower bound the MD probability in terms of the first three moments of the mismatched information density. This yields,
    $$P_{\text{MD}}(\gamma) \geq Q\left(-\frac{N\gamma-\sum_n c_n}{\sqrt{\sum_n v_n}}\right) - \frac{B}{\sqrt\Omega}.$$
Instead of optimizing over $\gamma$, we choose (sub-optimally)
    $$N\gamma = \sum_n c_n - \sqrt{\sum_n v_n} Q^{-1}{\left(P_e+\frac{2B}{\sqrt\Omega}\right)},$$
which results in the lowerbound:
    $$P_{\text{MD}}(\gamma) \geq P_e + \frac{B}{\sqrt\Omega}.$$
Substituting the above into the Verdú-Han bound \eqref{eq:VH_bound} and converting the units to bps/Hz (by noting $N$ FTN symbols are sent per $\Omega$ s$\cdot$Hz), we obtain:
    \begin{equation*}
    \begin{split}
        R & \leq  C_{\text{NA}} \!-\! \sqrt{\frac{V_{\text{NA}}}{\Omega}} \log_2(e) Q^{-1}\!{\left(P_e+\frac{2B}{\sqrt\Omega}\right)} \!-\! \frac{1}{\Omega} \log_2\!{\left(\frac{B}{\sqrt\Omega}\right)} \\
        & = C_{\text{NA}} \!-\! \sqrt{\frac{V_{\text{NA}}}{\Omega}} \log_2(e) Q^{-1}{(P_e)} + \frac{\log_2(\Omega)}{2\Omega} + \mathcal{O}\left(\frac{1}{\Omega}\right),
    \end{split}
    \end{equation*}
where in the last step, we used the Taylor series expansion on $Q^{-1}{(a+b)} \approx Q^{-1}(a) + \frac{Q^{-1'}(a)}{1!}b + \frac{Q^{-1''}(a)}{2!} b^2 + \cdots$ and collected all terms of order $\frac{1}{\Omega}$ into $\mathcal{O}(\frac{1}{\Omega})$. This proves Proposition \ref{prop:FTN_NA} and confirms that $R_\text{NA}$ is accurate up to $\mathcal{O}(\frac1 \Omega)$.
\end{proof}

\section{Approximation Of FTN MC and RCU Bounds} \label{Appendix:Approx}

Consider a linear combination $Y=\sum_{n=0}^{N-1}X_n$, where $X_n$ with PDF $f_{X_n}(x)$ are independent but not necessarily identically distributed. Let $K_Y(t)=\ln\mathbb{E}\{e^{tY}\}$ be the cumulative generating function (CGF) of $Y$ and $K'_Y(t)$ and $K''_Y(t)$ be the $1^{\text{st}}$ and $2^{\text{nd}}$ derivatives of $K_Y(t)$, respectively. Then, due to Corollary 1 of \cite{maya2016closed}, the logarithm of the CDF of $Y$ may be approximated as
    \begin{equation*} \begin{split}
    \ln P[Y \leq a] \approx & \ln Q\left(\sqrt{\hat{t}^2 K''_Y(\hat{t})}\right) \\ & + K_Y(\hat{t}) - \hat{t}K'_Y(\hat{t}) + \tfrac{\hat{t}^2}{2} K''_Y(\hat{t}),
    \end{split} \end{equation*}
where $\hat{t}\leq0$ satisfies $K'_Y(\hat{t})=a$. The aproximation is valid when $a \leq \mathbb{E}[Y]$ and $\int_{-\infty}^{\infty} \lvert x-K'_{X_n}(t) \rvert^3 f_{X_n}(x) e^{tx-K_{X_n}(t)} dx$ exist and are finite in a neighborhood of $\hat{t}$. 

When applied to the MC bound for FTN in Theorem \ref{thm:MC_FTN}, we get the following approximation for \eqref{eq:MC_FTN}:
    \begin{equation} \label{eq:MC_FTN_approx}
    \begin{split}
        R \approx -\frac{1}{\Omega \ln(2)} \bigg[\ln Q\left(\sqrt{\hat{t}^2 K''(\hat{t})}\right) + K(\hat{t}) \\ - \hat{t}K'(\hat{t}) + \tfrac{\hat{t}^2}{2} K''(\hat{t})\bigg],
    \end{split}
    \end{equation}
where $\hat{t}\leq0$ satisfies $K'(\hat{t})=N\lambda$ and
    \begin{equation*}
    \begin{split}
        &K(t)=\sum_{n=0}^{N-1} \left(\frac{\xi_n t}{\sigma_n^2-t}-\ln\left(1-\tfrac{1}{\sigma_n^2} t\right) \right),\\
        &K'(t)=\sum_{n=0}^{N-1} \left(\frac{1}{\sigma_n^2-t} \left(1+\frac{\xi_n}{1-\frac{1}{\sigma_n^2}t}\right)\right),\\
        &K''(t)=\sum_{n=0}^{N-1} \left(\frac{1}{\left(\sigma_n^2-t\right)^2} \left(1+\frac{2\xi_n}{1-\frac{1}{\sigma_n^2}t} \right)\right).
    \end{split}
    \end{equation*}
\noindent with $\xi_n \triangleq 1+\sigma_n^2$. 

This result follows from the moment generating function (MGF) of $\mathcal{X}^2(1,\nu)$ being $M(t)=\frac{1}{1-t} \exp{(\frac{t}{1-t}\nu)}$ \cite{turin1960characteristic}. The approximation \eqref{eq:MC_FTN_approx} is simple to compute due to its closed-form expression and $\hat{t}$ can be found using any root-finding algorithms. Computing $P_e$ \eqref{eq:Pe_MC}, on the other hand, does not need such approximation since the range of $P_e$ of interest are usually higher than $10^{-6}$. 

Similarly, applying the approximation to the RCU bound for FTN \eqref{eq:FTN_RCU_Pe} in Theorem \ref{thm:FTN_RCU}, we obtain
    \begin{equation} \label{eq:RCU_Pe_approx}
    \begin{split}
        P_e \approx \mathbb{E}_{\tilde{\mathbf{x}},\tilde{\mathbf{y}}} \Bigl[ \exp\Bigl(&\min \Bigl\{ 0, \ln(2^{\Omega R}-1) + \ln Q \left( \sqrt{\hat{t}_{\tilde{\mathbf{y}}}^2 K''(\hat{t}_{\tilde{\mathbf{y}}})}\right) \\
        &+ K(\hat{t}_{\tilde{\mathbf{y}}}) - \hat{t}_{\tilde{\mathbf{y}}} K'(\hat{t}_{\tilde{\mathbf{y}}}) + \tfrac{\hat{t}_{\tilde{\mathbf{y}}}^2}{2} K''(\hat{t}_{\tilde{\mathbf{y}}}) \Bigr\} \Bigr) \Bigr]
    \end{split}
    \end{equation}

\noindent where $\hat{t}_{\tilde{\mathbf{y}}}\leq0$ satisfies $K'(\hat{t}_{\tilde{\mathbf{y}}})=\mu(\tilde{\mathbf{x}},\tilde{\mathbf{y}})$, and $K(t)$, $K'(t)$, and $K''(t)$ are defined as above but using $\xi_n \triangleq |\tilde{y}_n|^2$.

For $\Omega R \gg 0$, we can use the approximation $\ln(2^{\Omega R}-1) \approx \Omega R \ln(2)$. We note that CGF $K(t)$ and its derivatives in the RCU approximation are all functions of $\tilde{\mathbf{y}}$, and $\hat{t}_{\tilde{\mathbf{y}}}$ must be found for each realization of $\tilde{\mathbf{y}}$. In our simulations, the expected value in \eqref{eq:RCU_Pe_approx} is numerically computed by taking a sample mean of at least $10^6$ realizations of $\tilde{\mathbf{x}}$ and $\tilde{\mathbf{y}}$.

\section{Proof of Theorem \ref{thm:FTN_RCU}} \label{Appendix:ProofRCU}
\begin{proof}
The (matched) information density between the input symbols $\tilde{\mathbf{x}}$ and the FTN channel output $\tilde{\mathbf{y}}$, when $\tilde{\mathbf{y}} \sim \mathcal{CN}(\mathbf{0},\mathbf{I}+\mathbf{D})$, may be expressed as
    \begin{equation} \label{eq:suminfodensity}
        i(\tilde{\mathbf{x}},\tilde{\mathbf{y}}) = \sum_{n=0}^{N-1} i(\tilde{x}_n;\tilde{y}_n),
    \end{equation}
where 
    \begin{equation} \label{eq:infodensity}
    \begin{split}
        i(\tilde{x}_n;\tilde{y}_n) &\triangleq \ln \left(\frac{p(\tilde{y}_n | \tilde{x}_n)}{p(\tilde{y}_n)}\right) \\ 
        &= \ln\left(1+\frac{1}{\sigma_n^2}\right) - \frac{|\tilde{y}_n-\tilde{x}_n|^2}{\sigma_n^2} + \frac{|\tilde{y}_n |^2}{1+\sigma_n^2}.
    \end{split}
    \end{equation}

\noindent To see this, first note that $\tilde{\mathbf{y}}|\tilde{\mathbf{x}} \sim \mathcal{CN}(\tilde{\mathbf{x}},\mathbf{D})$ and hence $p(\tilde{\mathbf{y}}|\tilde{\mathbf{x}}) = \prod_n p(\tilde{y}_n | \tilde{x}_n)$ and $p(\tilde{\mathbf{y}}) = \prod_n p(\tilde{y}_n)$, and \eqref{eq:suminfodensity} follows immediately. Substituting the corresponding complex normal PDFs and simplifying yields \eqref{eq:infodensity}. Then, $g(\tilde{\mathbf{x}},\tilde{\mathbf{y}})$ from \eqref{eq:g_xy} simplifies to:
    \begin{equation*}
    \begin{split}
        g(\tilde{\mathbf{x}},&\tilde{\mathbf{y}}) = P\left[i(\mathbf{w};\tilde{\mathbf{y}}) \geq i(\tilde{\mathbf{x}};\tilde{\mathbf{y}}) \middle| \tilde{\mathbf{x}},\tilde{\mathbf{y}}\right] \\ 
        &= P\left[\sum_{n=0}^{N-1} \frac{1}{\sigma_n^2} |w_n-\tilde{y}_n|^2 \leq \sum_{n=0}^{N-1} \frac{1}{\sigma_n^2} |\tilde{x}_n-\tilde{y}_n|^2 \middle| \tilde{\mathbf{x}},\tilde{\mathbf{y}}\right].
    \end{split}
    \end{equation*}

\noindent Finally, $|w_n-\tilde{y}_n|^2 \sim \mathcal{X}^2(1,|\tilde{y}_n|^2)$ when $\mathbf{w} \sim p_{\tilde{\mathbf{x}}}=\mathcal{CN}(\mathbf{0},\mathbf{I})$. This completes the proof of Theorem \ref{thm:FTN_RCU}.
\end{proof}

\section{PSWF: Definitions, properties, and benchmarks} \label{Appendix:PSWF}

Consider the time interval $|t| \leq T_x/2$ and bandwidth $|f| \leq W$. The corresponding TBP is $\Omega=2WT_x$. Prolate Spheroidal Wave Functions (PSWFs) are eigenfunctions of the integral equation:
    $$\mu_{\Omega,n} \psi_{\Omega,n}(t) = \int_{-T_x/2}^{T_x/2} 2W\text{sinc}(2W(t-s)) \psi_{\Omega,n} (s)ds,$$
where $1\geq\mu_{\Omega,0}>\mu_{\Omega,1}>\cdots>0$ are the associated eigenvalues. While ${\psi_{\Omega,n}(t)}$ lack closed-form expressions, many of their properties are well known \cite{slepian1961prolate}, \cite{slepian1983some}. They are real-valued, bandlimited, orthonormal on the real line, and complete in the space of bandlimited functions. Furthermore, the energy concentration in $|t| \leq T_x/2$ is equal to the eigenvalues $\mu_{\Omega,n}$. $\psi_{\Omega,0}(t)$ has the highest energy concentration in $|t| \leq T_x/2$ among all bandlimited signals, $\psi_{\Omega,1}(t)$ has the highest energy among all signals orthogonal to $\psi_{\Omega,0}(t)$, and so on. 

Let $\phi_{\Omega,n}(t) \triangleq \frac{\mathcal{D}\psi_{\Omega,n}(t)}{\sqrt{\mu_{\Omega,n}}}$ denote normalized and time-truncated PSWFs with the time support $|t| \leq \frac{T_x}2$ and a unit energy. These are orthonormal and complete in the space of time-limited functions in $|t| \leq T_x/2$. Furthermore, its energy concentration within the band $|f|\leq W$ is also $\mu_{\Omega,n}$, i.e., $\int_{-W}^{W}{|\hat{\phi}_{\Omega,n}(f)|^2 df} = \mu_{\Omega,n}$, where $\hat{\phi}_{\Omega,n}(f)$ is the Fourier transform of $\phi_{\Omega,n}(t)$.

Due to their completeness, any time-limited signal with a TBP of $\Omega$ may be written as a linear combination of the normalized and truncated PSWFs: i.e., $x(t) = \sum_{n=0}^{\infty} a_n \phi_{\Omega,n}(t)$. Using the orthonormality of the PSWFs, the OOB constraint may be expressed as $\sum_{n=0}^{\infty} |a_n|^2 (1-\mu_{\Omega,n}) \leq \epsilon_W \sum_{n=0}^{\infty} |a_n|^2$, from which we see that $\{a_n\}$ should approach $0$ for large $n$ due to the eigenvalues decreasing in $n$. It is also known that, for some $\eta>0$, $\mu_{\Omega,n}$ decreases exponentially towards zero for all $n>\Omega-\eta$ when $\Omega$ is large, which forces $a_n \to 0$ exponentially fast for $n>\Omega-\eta$ in the high TBP regime. 

The above observations motivate us to consider transmitting $N$ data symbols using the normalized and truncated PSWFs:
    $$x_{\text{PSWF}}(t)=\sum_{n=0}^{N-1} x_n \phi_{\Omega,n}(t),$$
where $N$ is chosen to be the largest possible integer while the OOB constraint is met. Since $\{\phi_{\Omega,n}(t)\}$ are orthonormal, SNR experienced by the individual symbols are equal. We thus choose \textit{i.i.d.} $x_n \sim \mathcal{CN}(0,\frac{PT_x}{N})$ with uniform power to satisfy the power constraint. The corresponding OOB constraint is: 
    $$\frac{1}{N} \sum_{n=0}^{N-1} (1-\mu_{\Omega,n}) \leq \epsilon_W,$$ 
and $N$ is determined by incrementing $N$ above until the constraint is no longer satisfied. The corresponding NA of MCCR is given by \eqref{eq:R_NA} with the $n$-th channel SNR $\sigma_n^2 = \left(\rho\frac{\Omega}{N}\right)^{-1}$ and $N$ as determined by the procedure explained above. 

We note that this PSWF benchmark is suboptimal due to lacking symbol power optimization. Nevertheless, it provides a simple expression (without requiring any optimization) and a fair benchmark for the considered \textit{u.i.d.} FTN signaling without symbol precoding. 

The optimal PSWF benchmark (optimality in the sense of maximizing the capacity) is obtained by solving the following constrained optimization problem:
    \begin{equation*} \begin{aligned}
        \underset{\{P_n\geq0\},\forall n}{\arg\max} &\left\{\sum_{n=0}^\infty \log_2\left(1+\frac{P_n}{N_0}\right) \right\} 
    \end{aligned} \end{equation*}
    \begin{equation*} \begin{aligned}
        \text{s.t.}  \quad &\frac{1}{PT_x} \sum_{n=0}^{\infty} P_n(1-\mu_{\Omega,n}) \leq \epsilon_W \text{ (OOB)} \\
        & \frac{1}{PT_x} \sum_{n=0}^{\infty} P_n \leq 1 \text{ (power constraint)}
    \end{aligned}  \end{equation*}
assuming $a_n \sim \mathcal{CN}(0,P_n)$ and independent in $n$. It is easy to see that the maximum is attained when both constraints are met with equalities (i.e., using the most available power and most allowed OOB). Using the method of Lagrange multiplier, the optimal $P_n$ is given by the waterfilling:
    $$P_n = \max\left\{0,\frac{PT_x}{\theta_1 (1-\mu_{\Omega,n} )+\theta_2}-N_0 \right\},$$
where $\theta_1>0$ and $\theta_2>0$ are chosen such that the constraints are satisfied with equalities. The corresponding NA of MCCR is given by \eqref{eq:R_NA} with $\sigma_n^2=(P_n/N_0)^{-1}$ with $P_n$ determined by the optimization problem above. 

In the case of bandlimited signal with OOI constraint, the transmitted signal in the baseband may be expressed as a linear combination of the bandlimited PSWFs (non-time truncated):
    $$x(t) = \sum_{n=0}^{\infty} a_n \psi_{\Omega,n}(t)$$
due to the PSWFs being complete in the space of bandlimited functions. In addition, due to the orthonormality of PSWFs, the OOI constraint may be expressed as $\sum_{n=0}^{\infty} |a_n|^2 (1-\mu_{\Omega,n}) \leq \epsilon_T \sum_{n=0}^{\infty} |a_n|^2$, which is identical to the OOB constraint with $\epsilon_W$ replaced by $\epsilon_T$. In other words, the PSWF benchmarks with OOB or OOI constraints are exactly the same when $\epsilon_W=\epsilon_T$, even though the transmitted signals are different. 

It should be noted that, although PSWFs yield the optimal transmission scheme for a given TBP, they are difficult to use in practice due to 1) lack of closed-form expressions making them difficult to generate with high precision, 2) number of functions that needs to be computed by both the modulator and the demodulator increases with TBP, and 3) the entire set of functions must be recomputed if TBP changes. The PSWF benchmarks should be understood as theoretical benchmarks for designing FTN systems that are otherwise difficult to approach with the conventional Nyquist systems. 

\bibliographystyle{IEEEtran}
\bibliography{IEEEabrv, references}

\end{document}

%% file: notation_new.tex
\def\nba{{\mathbf{a}}}
\def\nbb{{\mathbf{b}}}
\def\nbc{{\mathbf{c}}}
\def\nbd{{\mathbf{d}}}
\def\nbe{{\mathbf{e}}}
\def\nbf{{\mathbf{f}}}
\def\nbg{{\mathbf{g}}}
\def\nbh{{\mathbf{h}}}
\def\nbi{{\mathbf{i}}}
\def\nbj{{\mathbf{j}}}
\def\nbk{{\mathbf{k}}}
\def\nbl{{\mathbf{l}}}
\def\nbm{{\mathbf{m}}}
\def\nbn{{\mathbf{n}}}
\def\nbo{{\mathbf{o}}}
\def\nbp{{\mathbf{p}}}
\def\nbq{{\mathbf{q}}}
\def\nbr{{\mathbf{r}}}
\def\nbs{{\mathbf{s}}}
\def\nbt{{\mathbf{t}}}
\def\nbu{{\mathbf{u}}}
\def\nbv{{\mathbf{v}}}
\def\nbw{{\mathbf{w}}}
\def\nbx{{\mathbf{x}}}
\def\nby{{\mathbf{y}}}
\def\nbz{{\mathbf{z}}}
\def\nb0{{\mathbf{0}}}
\def\nb1{{\mathbf{1}}}

\def\nbA{{\mathbf{A}}}
\def\nbB{{\mathbf{B}}}
\def\nbC{{\mathbf{C}}}
\def\nbD{{\mathbf{D}}}
\def\nbE{{\mathbf{E}}}
\def\nbF{{\mathbf{F}}}
\def\nbG{{\mathbf{G}}}
\def\nbH{{\mathbf{H}}}
\def\nbI{{\mathbf{I}}}
\def\nbJ{{\mathbf{J}}}
\def\nbK{{\mathbf{K}}}
\def\nbL{{\mathbf{L}}}
\def\nbM{{\mathbf{M}}}
\def\nbN{{\mathbf{N}}}
\def\nbO{{\mathbf{O}}}
\def\nbP{{\mathbf{P}}}
\def\nbQ{{\mathbf{Q}}}
\def\nbR{{\mathbf{R}}}
\def\nbS{{\mathbf{S}}}
\def\nbT{{\mathbf{T}}}
\def\nbU{{\mathbf{U}}}
\def\nbV{{\mathbf{V}}}
\def\nbW{{\mathbf{W}}}
\def\nbX{{\mathbf{X}}}
\def\nbY{{\mathbf{Y}}}
\def\nbZ{{\mathbf{Z}}}

\def\nbtheta{\boldsymbol{\theta}}
\def\nbphi{\boldsymbol{\phi}}

\def\nsfa{{\mathsf{a}}}
\def\nsfb{{\mathsf{b}}}
\def\nsfc{{\mathsf{c}}}
\def\nsfd{{\mathsf{d}}}
\def\nsfe{{\mathbf{e}}}
\def\nsff{{\mathbf{f}}}
\def\nsfg{{\mathbf{g}}}
\def\nsfh{{\mathbf{h}}}
\def\nsfi{{\mathbf{i}}}
\def\nsfj{{\mathbf{j}}}
\def\nsfk{{\mathbf{k}}}
\def\nsfl{{\mathbf{l}}}
\def\nsfm{{\mathbf{m}}}
\def\nsfn{{\mathbf{n}}}
\def\nsfo{{\mathbf{o}}}
\def\nsfp{{\mathbf{p}}}
\def\nsfq{{\mathbf{q}}}
\def\nsfr{{\mathbf{r}}}
\def\nsfs{{\mathbf{s}}}
\def\nsft{{\mathbf{t}}}
\def\nsfu{{\mathbf{u}}}
\def\nsfv{{\mathbf{v}}}
\def\nsfw{{\mathbf{w}}}
\def\nsfx{{\mathbf{x}}}
\def\nsfy{{\mathbf{y}}}
\def\nsfz{{\mathbf{z}}}

\def\nsfA{{\mathsf{A}}}
\def\nsfB{{\mathsf{B}}}
\def\nsfC{{\mathsf{C}}}
\def\nsfD{{\mathsf{D}}}
\def\nsfE{{\mathsf{E}}}
\def\nsfF{{\mathsf{F}}}
\def\nsfG{{\mathsf{G}}}
\def\nsfH{{\mathsf{H}}}
\def\nsfI{{\mathsf{I}}}
\def\nsfJ{{\mathsf{J}}}
\def\nsfK{{\mathsf{K}}}
\def\nsfL{{\mathsf{L}}}
\def\nsfM{{\mathsf{M}}}
\def\nsfN{{\mathsf{N}}}
\def\nsfO{{\mathsf{O}}}
\def\nsfP{{\mathsf{P}}}
\def\nsfQ{{\mathsf{Q}}}
\def\nsfR{{\mathsf{R}}}
\def\nsfS{{\mathsf{S}}}
\def\nsfT{{\mathsf{T}}}
\def\nsfU{{\mathsf{U}}}
\def\nsfV{{\mathsf{V}}}
\def\nsfW{{\mathsf{W}}}
\def\nsfX{{\mathsf{X}}}
\def\nsfY{{\mathsf{Y}}}
\def\nsfZ{{\mathsf{Z}}}

\def\nsfTheta{\mathsf{\Theta}}
\def\nbphi{\boldsymbol{\phi}}

\def\ncalA{{\mathcal{A}}}
\def\ncalB{{\mathcal{B}}}
\def\ncalC{{\mathcal{C}}}
\def\ncalD{{\mathcal{D}}}
\def\ncalE{{\mathcal{E}}}
\def\ncalF{{\mathcal{F}}}
\def\ncalG{{\mathcal{G}}}
\def\ncalH{{\mathcal{H}}}
\def\ncalI{{\mathcal{I}}}
\def\ncalJ{{\mathcal{J}}}
\def\ncalK{{\mathcal{K}}}
\def\ncalL{{\mathcal{L}}}
\def\ncalM{{\mathcal{M}}}
\def\ncalN{{\mathcal{N}}}
\def\ncalO{{\mathcal{O}}}
\def\ncalP{{\mathcal{P}}}
\def\ncalQ{{\mathcal{Q}}}
\def\ncalR{{\mathcal{R}}}
\def\ncalS{{\mathcal{S}}}
\def\ncalT{{\mathcal{T}}}
\def\ncalU{{\mathcal{U}}}
\def\ncalV{{\mathcal{V}}}
\def\ncalW{{\mathcal{W}}}
\def\ncalX{{\mathcal{X}}}
\def\ncalY{{\mathcal{Y}}}
\def\ncalZ{{\mathcal{Z}}}

\def\nbbA{{\mathbb{A}}}
\def\nbbB{{\mathbb{B}}}
\def\nbbC{{\mathbb{C}}}
\def\nbbD{{\mathbb{D}}}
\def\nbbE{{\mathbb{E}}}
\def\nbbF{{\mathbb{F}}}
\def\nbbG{{\mathbb{G}}}
\def\nbbH{{\mathbb{H}}}
\def\nbbI{{\mathbb{I}}}
\def\nbbJ{{\mathbb{J}}}
\def\nbbK{{\mathbb{K}}}
\def\nbbL{{\mathbb{L}}}
\def\nbbM{{\mathbb{M}}}
\def\nbbN{{\mathbb{N}}}
\def\nbbO{{\mathbb{O}}}
\def\nbbP{{\mathbb{P}}}
\def\nbbQ{{\mathbb{Q}}}
\def\nbbR{{\mathbb{R}}}
\def\nbbS{{\mathbb{S}}}
\def\nbbT{{\mathbb{T}}}
\def\nbbU{{\mathbb{U}}}
\def\nbbV{{\mathbb{V}}}
\def\nbbW{{\mathbb{W}}}
\def\nbbX{{\mathbb{X}}}
\def\nbbY{{\mathbb{Y}}}
\def\nbbZ{{\mathbb{Z}}}

\def\nfrakR{{\mathfrak{R}}}

\def\nrma{{\rm a}}
\def\nrmb{{\rm b}}
\def\nrmc{{\rm c}}
\def\nrmd{{\rm d}}
\def\nrme{{\rm e}}
\def\nrmf{{\rm f}}
\def\nrmg{{\rm g}}
\def\nrmh{{\rm h}}
\def\nrmi{{\rm i}}
\def\nrmj{{\rm j}}
\def\nrmk{{\rm k}}
\def\nrml{{\rm l}}
\def\nrmm{{\rm m}}
\def\nrmn{{\rm n}}
\def\nrmo{{\rm o}}
\def\nrmp{{\rm p}}
\def\nrmq{{\rm q}}
\def\nrmr{{\rm r}}
\def\nrms{{\rm s}}
\def\nrmt{{\rm t}}
\def\nrmu{{\rm u}}
\def\nrmv{{\rm v}}
\def\nrmw{{\rm w}}
\def\nrmx{{\rm x}}
\def\nrmy{{\rm y}}
\def\nrmz{{\rm z}}

\def\nbydef{:=}
\def\nborel{\ncalB(\nbbR)}
\def\nboreld{\ncalB(\nbbR^d)}
\def\sinc{{\rm sinc}}

\newtheorem{thm}{Theorem}
\newtheorem{lem}{Lemma}
\newtheorem{ndef}{Definition}
\newtheorem{nrem}{Remark}
\newtheorem{prop}{Proposition}
\newtheorem{cor}{Corollary}
\newtheorem{eg}{Example}
\newtheorem{assumption}{Assumption}
\newtheorem{approximation}{Approximation}

%% file: figNparallChan.tex
\tikzset{every picture/.style={line width=0.75pt}} 

\begin{tikzpicture}[x=0.75pt,y=0.75pt,yscale=-1,xscale=1]

\draw    (44,32) -- (108.82,32) ;
\draw [shift={(110.82,32)}, rotate = 180] [color={rgb, 255:red, 0; green, 0; blue, 0 }  ][line width=0.75]    (10.93,-3.29) .. controls (6.95,-1.4) and (3.31,-0.3) .. (0,0) .. controls (3.31,0.3) and (6.95,1.4) .. (10.93,3.29)   ;
\draw   (110.82,32) .. controls (110.82,27.38) and (114.56,23.64) .. (119.18,23.64) .. controls (123.8,23.64) and (127.55,27.38) .. (127.55,32) .. controls (127.55,36.62) and (123.8,40.36) .. (119.18,40.36) .. controls (114.56,40.36) and (110.82,36.62) .. (110.82,32) -- cycle ;
\draw    (119.18,40.36) -- (119.18,23.64) ;
\draw    (110.82,32) -- (127.55,32) ;

\draw    (127.55,32) -- (237.18,32) ;
\draw [shift={(239.18,32)}, rotate = 180] [color={rgb, 255:red, 0; green, 0; blue, 0 }  ][line width=0.75]    (10.93,-3.29) .. controls (6.95,-1.4) and (3.31,-0.3) .. (0,0) .. controls (3.31,0.3) and (6.95,1.4) .. (10.93,3.29)   ;
\draw    (119.18,57.27) -- (119.18,42.36) ;
\draw [shift={(119.18,40.36)}, rotate = 90] [color={rgb, 255:red, 0; green, 0; blue, 0 }  ][line width=0.75]    (10.93,-3.29) .. controls (6.95,-1.4) and (3.31,-0.3) .. (0,0) .. controls (3.31,0.3) and (6.95,1.4) .. (10.93,3.29)   ;
\draw    (136.82,57.27) -- (119.18,57.27) ;
\draw    (45,86) -- (109.82,86) ;
\draw [shift={(111.82,86)}, rotate = 180] [color={rgb, 255:red, 0; green, 0; blue, 0 }  ][line width=0.75]    (10.93,-3.29) .. controls (6.95,-1.4) and (3.31,-0.3) .. (0,0) .. controls (3.31,0.3) and (6.95,1.4) .. (10.93,3.29)   ;
\draw   (111.82,86) .. controls (111.82,81.38) and (115.56,77.64) .. (120.18,77.64) .. controls (124.8,77.64) and (128.55,81.38) .. (128.55,86) .. controls (128.55,90.62) and (124.8,94.36) .. (120.18,94.36) .. controls (115.56,94.36) and (111.82,90.62) .. (111.82,86) -- cycle ;
\draw    (120.18,94.36) -- (120.18,77.64) ;
\draw    (111.82,86) -- (128.55,86) ;

\draw    (128.55,86) -- (236.18,86) ;
\draw [shift={(238.18,86)}, rotate = 180] [color={rgb, 255:red, 0; green, 0; blue, 0 }  ][line width=0.75]    (10.93,-3.29) .. controls (6.95,-1.4) and (3.31,-0.3) .. (0,0) .. controls (3.31,0.3) and (6.95,1.4) .. (10.93,3.29)   ;
\draw    (120.18,111.27) -- (120.18,96.36) ;
\draw [shift={(120.18,94.36)}, rotate = 90] [color={rgb, 255:red, 0; green, 0; blue, 0 }  ][line width=0.75]    (10.93,-3.29) .. controls (6.95,-1.4) and (3.31,-0.3) .. (0,0) .. controls (3.31,0.3) and (6.95,1.4) .. (10.93,3.29)   ;
\draw    (137.82,111.27) -- (120.18,111.27) ;
\draw   (33.08,120.63) .. controls (33.08,120.1) and (33.51,119.67) .. (34.05,119.67) .. controls (34.58,119.67) and (35.02,120.1) .. (35.02,120.63) .. controls (35.02,121.17) and (34.58,121.6) .. (34.05,121.6) .. controls (33.51,121.6) and (33.08,121.17) .. (33.08,120.63) -- cycle ;
\draw   (33.08,126.63) .. controls (33.08,126.1) and (33.51,125.67) .. (34.05,125.67) .. controls (34.58,125.67) and (35.02,126.1) .. (35.02,126.63) .. controls (35.02,127.17) and (34.58,127.6) .. (34.05,127.6) .. controls (33.51,127.6) and (33.08,127.17) .. (33.08,126.63) -- cycle ;
\draw   (33.08,133.63) .. controls (33.08,133.1) and (33.51,132.67) .. (34.05,132.67) .. controls (34.58,132.67) and (35.02,133.1) .. (35.02,133.63) .. controls (35.02,134.17) and (34.58,134.6) .. (34.05,134.6) .. controls (33.51,134.6) and (33.08,134.17) .. (33.08,133.63) -- cycle ;

\draw    (45.67,155.55) -- (110.48,155.55) ;
\draw [shift={(112.48,155.55)}, rotate = 180] [color={rgb, 255:red, 0; green, 0; blue, 0 }  ][line width=0.75]    (10.93,-3.29) .. controls (6.95,-1.4) and (3.31,-0.3) .. (0,0) .. controls (3.31,0.3) and (6.95,1.4) .. (10.93,3.29)   ;
\draw   (112.48,155.55) .. controls (112.48,150.93) and (116.23,147.18) .. (120.85,147.18) .. controls (125.47,147.18) and (129.21,150.93) .. (129.21,155.55) .. controls (129.21,160.16) and (125.47,163.91) .. (120.85,163.91) .. controls (116.23,163.91) and (112.48,160.16) .. (112.48,155.55) -- cycle ;
\draw    (120.85,163.91) -- (120.85,147.18) ;
\draw    (112.48,155.55) -- (129.21,155.55) ;

\draw    (129.21,155.55) -- (236.85,155.55) ;
\draw [shift={(238.85,155.55)}, rotate = 180] [color={rgb, 255:red, 0; green, 0; blue, 0 }  ][line width=0.75]    (10.93,-3.29) .. controls (6.95,-1.4) and (3.31,-0.3) .. (0,0) .. controls (3.31,0.3) and (6.95,1.4) .. (10.93,3.29)   ;
\draw    (120.85,180.82) -- (120.85,165.91) ;
\draw [shift={(120.85,163.91)}, rotate = 90] [color={rgb, 255:red, 0; green, 0; blue, 0 }  ][line width=0.75]    (10.93,-3.29) .. controls (6.95,-1.4) and (3.31,-0.3) .. (0,0) .. controls (3.31,0.3) and (6.95,1.4) .. (10.93,3.29)   ;
\draw    (138.48,180.82) -- (120.85,180.82) ;
\draw   (119.83,122.63) .. controls (119.83,122.1) and (120.26,121.67) .. (120.8,121.67) .. controls (121.33,121.67) and (121.77,122.1) .. (121.77,122.63) .. controls (121.77,123.17) and (121.33,123.6) .. (120.8,123.6) .. controls (120.26,123.6) and (119.83,123.17) .. (119.83,122.63) -- cycle ;
\draw   (119.83,128.63) .. controls (119.83,128.1) and (120.26,127.67) .. (120.8,127.67) .. controls (121.33,127.67) and (121.77,128.1) .. (121.77,128.63) .. controls (121.77,129.17) and (121.33,129.6) .. (120.8,129.6) .. controls (120.26,129.6) and (119.83,129.17) .. (119.83,128.63) -- cycle ;
\draw   (119.83,135.63) .. controls (119.83,135.1) and (120.26,134.67) .. (120.8,134.67) .. controls (121.33,134.67) and (121.77,135.1) .. (121.77,135.63) .. controls (121.77,136.17) and (121.33,136.6) .. (120.8,136.6) .. controls (120.26,136.6) and (119.83,136.17) .. (119.83,135.63) -- cycle ;

\draw (140,48.4) node [anchor=north west][inner sep=0.75pt]  [font=\footnotesize]  {$\tilde{z}_{0} \sim \mathcal{CN}\left( 0,\sigma _{0}^{2}\right)$};
\draw (20,23.4) node [anchor=north west][inner sep=0.75pt]  [font=\footnotesize]  {$\tilde{x}_{0}$};
\draw (242,22.4) node [anchor=north west][inner sep=0.75pt]  [font=\footnotesize]  {$\tilde{y}_{0}$};
\draw (142,102.4) node [anchor=north west][inner sep=0.75pt]  [font=\footnotesize]  {$\tilde{z}_{1} \sim \mathcal{CN}\left( 0,\sigma _{1}^{2}\right)$};
\draw (20,76.4) node [anchor=north west][inner sep=0.75pt]  [font=\footnotesize]  {$\tilde{x}_{1}$};
\draw (243,76.4) node [anchor=north west][inner sep=0.75pt]  [font=\footnotesize]  {$\tilde{y}_{1}$};
\draw (142.67,171.95) node [anchor=north west][inner sep=0.75pt]  [font=\footnotesize]  {$\tilde{z}_{N-1} \sim \mathcal{CN}\left( 0,\sigma _{N-1}^{2}\right)$};
\draw (12,147.28) node [anchor=north west][inner sep=0.75pt]  [font=\footnotesize]  {$\tilde{x}_{N-1}$};
\draw (243.67,145.95) node [anchor=north west][inner sep=0.75pt]  [font=\footnotesize]  {$\tilde{y}_{N-1}$};

\end{tikzpicture}

%% file: references.bib
@IEEEtranBSTCTL{IEEEexample:BSTcontrol,
CTLuse_forced_etal = "yes",
CTLmax_names_forced_etal = "4",
CTLnames_show_etal = "1"
}

@article{anderson2013faster,
  title={Faster-than-{N}yquist signaling},
  author={Anderson, John B. and Rusek, Fredrik and {\"O}wall, Viktor},
  journal={Proceedings of the IEEE},
  volume={101},
  number={8},
  pages={1817--1830},
  year={2013}, month=aug,
  publisher={IEEE}
}

@article{fan2017faster,
  title={Faster-than-{N}yquist signaling: An overview},
  author={Fan, Jiancun and Guo, Shengjie and Zhou, Xiangwei and Ren, Yajie and Li, Geoffrey Ye and Chen, Xi},
  journal={IEEE Access},
  volume={5},
  pages={1925--1940},
  year={2017},
  publisher={IEEE}
}

@article{ishihara2021evolution,
  title={The evolution of faster-than-{N}yquist signaling},
  author={Ishihara, Takumi and Sugiura, Shinya and Hanzo, Lajos},
  journal={IEEE Access},
  volume={9},
  pages={86535--86564},
  year={2021},
  publisher={IEEE}
}

@article{rusek2009constrained,
  title={Constrained capacities for faster-than-{N}yquist signaling},
  author={Rusek, Fredrik and Anderson, John B.},
  journal={IEEE Transactions on Information Theory},
  volume={55},
  number={2},
  pages={764--775},
  year={2009}, month=feb,
  publisher={IEEE}
}

@article{liveris2003exploiting,
  title={Exploiting faster-than-{N}yquist signaling},
  author={Liveris, Angelos D and Georghiades, Costas N},
  journal={IEEE Transactions on Communications},
  volume={51},
  number={9},
  pages={1502--1511},
  year={2003}, month=sep,
  publisher={IEEE}
}

@article{prlja2012reduced,
  title={Reduced-complexity receivers for strongly narrowband intersymbol interference introduced by faster-than-{N}yquist signaling},
  author={Prlja, Adnan and Anderson, John B.},
  journal={IEEE Transactions on Communications},
  volume={60},
  number={9},
  pages={2591--2601},
  year={2012}, month=sep,
  publisher={IEEE}
}

@article{sugiura2015frequency,
  title={Frequency-domain-equalization-aided iterative detection of faster-than-{N}yquist signaling},
  author={Sugiura, Shinya and Hanzo, Lajos},
  journal={IEEE Transactions on Vehicular Technology},
  volume={64},
  number={5},
  pages={2122--2128},
  year={2015}, month=may,
  publisher={IEEE}
}

@article{kim2016faster,
  title={Faster-than-{N}yquist broadcasting in {G}aussian channels: Achievable rate regions and coding},
  author={Kim, Yong Jin Daniel and Bajcsy, Jan and Vargas, David},
  journal={IEEE Transactions on Communications},
  volume={64},
  number={3},
  pages={1016--1030},
  year={2016}, month=mar,
  publisher={IEEE}
}

@article{shirvanimoghaddam2018short,
  title={Short block-length codes for ultra-reliable low latency communications},
  author={Shirvanimoghaddam, Mahyar and Mohammadi, Mohammad Sadegh and Abbas, Rana and Minja, Aleksandar and Yue, Chentao and Matuz, Balazs and Han, Guojun and Lin, Zihuai and Liu, Wanchun and Li, Yonghui and others},
  journal={IEEE Communications Magazine},
  volume={57},
  number={2},
  pages={130--137},
  year={2019}, month=feb,
  publisher={IEEE}
}

@article{mohammadkarimi2020channel,
  title={Channel coding rate for finite blocklength faster-than-{N}yquist signaling},
  author={Mohammadkarimi, Mostafa and Schober, Robert and Wong, Vincent W. S.},
  journal={IEEE Communications Letters},
  volume={25},
  number={1},
  pages={64--68},
  year={2020}, month=jan,
  publisher={IEEE}
}

@inproceedings{zhang2025maximum,
  title={Maximum channel coding rate of finite block length {MIMO} faster-than-{N}yquist signaling},
  author={Zhang, Zichao and Yuksel, Melda and Yanikomeroglu, Halim and Ng, Benjamin K. and Lam, Chan--Tong},
  booktitle={Proc. IEEE Wireless Communications and Networking Conference (WCNC)},
  pages={1--6},
  year={2025}, month=mar, address={Milan, Italy},
}

@article{wyner1966capacity,
  title={The capacity of the band-limited {G}aussian channel},
  author={Wyner, Aaron Daniel},
  journal={The Bell System Technical Journal},
  volume={45},
  number={3},
  pages={359--395},
  year={1966}, month=mar,
  publisher={Wiley Online Library}
}

@book{gallager1968information,
  title={Information theory and reliable communication},
  author={Gallager, Robert G},
  year={1968},
  publisher={New York : Wiley}
}

@article{jaffal2020time,
  title={Time-limited codewords over band-limited channels: {D}ata rates and the dimension of the {W}-{T} space},
  author={Jaffal, Youssef and Abou-Faycal, Ibrahim},
  journal={Entropy},
  volume={22},
  number={9},
  pages={924},
  year={2020}, 
  publisher={MDPI}
}

@article{ishihara2021eigendecomposition,
  title={Eigendecomposition-precoded faster-than-{N}yquist signaling with optimal power allocation in frequency-selective fading channels},
  author={Ishihara, Takumi and Sugiura, Shinya},
  journal={IEEE Transactions on Wireless Communications},
  volume={21},
  number={3},
  pages={1681--1693},
  year={2022}, month=mar,
  publisher={IEEE}
}

@inproceedings{kim2010spectrum,
  title={On spectrum broadening of pre-coded faster-than-{N}yquist signaling},
  author={Kim, Yong Jin Daniel and Bajcsy, Jan},
  booktitle={Proc. IEEE 72nd Vehicular Technology Conference (VTC)-Fall},
  pages={1--5},
  year={2010}, month=sep,
  address={Ottawa, ON, Canada},
}

@inproceedings{kim2016properties,
  title={Properties of faster-than-{N}yquist channel matrices and folded-spectrum, and their applications},
  author={Kim, Yong Jin Daniel},
  booktitle={Proc. IEEE Wireless Communications and Networking Conference (WCNC)},
  pages={1982--1988},
  year={2016}, month=apr,
  address={Doha, Qatar},
}

@inproceedings{gattami2015time,
  title={Time localization and capacity of faster-than-{N}yquist signaling},
  author={Gattami, Ather and Ringh, Emil and Karlsson, Johan},
  booktitle={Proc. IEEE Global Communications Conference (GLOBECOM)},
  pages={1--7},
  year={2015}, month=dec,
  address={San Diego, CA, USA},
}

@article{scarlett2016dispersion,
  title={The dispersion of nearest-neighbor decoding for additive non-{G}aussian channels},
  author={Scarlett, Jonathan and Tan, Vincent Y. F. and Durisi, Giuseppe},
  journal={IEEE Transactions on Information Theory},
  volume={63},
  number={1},
  pages={81--92},
  year={2017}, month=jan,
  publisher={IEEE}
}

@book{cover1999elements,
  title={Elements of information theory},
  author={Cover, Thomas M. and Thomas, Joy A.},
  year={1991}, 
  publisher={New York : Wiley-Interscience}
}

@article{erseghe2016coding,
  title={Coding in the finite-blocklength regime: Bounds based on {L}aplace integrals and their asymptotic approximations},
  author={Erseghe, Tomaso},
  journal={IEEE Transactions on Information Theory},
  volume={62},
  number={12},
  pages={6854--6883},
  year={2016}, month=dec,
  publisher={IEEE}
}

@article{gursoy2013throughput,
  title={Throughput analysis of buffer-constrained wireless systems in the finite blocklength regime},
  author={Gursoy, M. Cenk},
  journal={EURASIP Journal on Wireless Communications and Networking},
  volume={2013}, number={290},
  pages={1--13},
  year={2013},
  publisher={Springer},
}

@article{polyanskiy2010channel,
  title={Channel coding rate in the finite blocklength regime},
  author={Polyanskiy, Yury and Poor, H Vincent and Verd{\'u}, Sergio},
  journal={IEEE Transactions on Information Theory},
  volume={56},
  number={5},
  pages={2307--2359},
  year={2010}, month=may,
  publisher={IEEE}
}

@inproceedings{font2018saddlepoint,
  title={Saddlepoint approximations of lower and upper bounds to the error probability in channel coding},
  author={Font-Segura, Josep and Vazquez-Vilar, Gonzalo and Martinez, Alfonso and i F{\`a}bregas, Albert Guill{\'e}n and Lancho, Alejandro},
  booktitle={Proc. 52nd Annual Conference on Information Sciences and Systems (CISS)},
  pages={1--6},
  year={2018}, month=mar, 
  address={Princeton, NJ, USA},
}

@article{maya2016closed,
  title={A closed-form approximation for the {CDF} of the sum of independent random variables},
  author={Maya, Juan Augusto and Vega, Leonardo Rey and Galarza, Cecilia G.},
  journal={IEEE Signal Processing Letters},
  volume={24},
  number={1},
  pages={121--125},
  year={2017}, month=jan,
  publisher={IEEE}
}

@misc{durisi2020lecture,
title = {Lecture notes: Transmitting short packets over wireless channels -- an information theoretic perspective},
author={Durisi, G. and Lancho, A.},
url={https://gdurisi.github.io/fbl-notes/index.html},
year = {2020},
}

@article{slepian1983some,
  title={Some comments on {F}ourier analysis, uncertainty and modeling},
  author={Slepian, David},
  journal={SIAM review},
  volume={25},
  number={3},
  pages={379--393},
  year={1983}, month=jul,
  publisher={SIAM}
}

@inproceedings{le2014practical,
  title={On the practical benefits of faster-than-{N}yquist signaling},
  author={Le, Chung and Schellmann, Malte and Fuhrwerk, Martin and Peissig, J{\"u}rgen},
  booktitle={Proc. International Conference on Advanced Technologies for Communications (ATC)},
  pages={208--213},
  year={2014}, month=oct,
  address={Hanoi, Vietnam}
}

@inproceedings{rusek2008faster,
  title={Faster-than-{N}yquist modulation based on short finite pulses},
  author={Rusek, Fredrik and Prlja, Adnan and Kapetanovi{\'c}, Dzevdan and Anderson, John B.},
  booktitle={Proc. Nordic Radio Science and Communication Conference (RVK)},
  pages={166--170}, month=jun,
  year={2008}, 
  address={V\"axj\"o, Sweden}
}

@article{foschini1984contrasting,
  title={Contrasting performance of faster binary signaling with {QAM}},
  author={Foschini, Gerard J.},
  journal={AT\&T Bell Laboratories Technical Journal},
  volume={63},
  number={8},
  pages={1419--1445}, month=oct,
  year={1984},
}

@inproceedings{anderson2007optimal,
  title={Optimal side lobes under linear and faster-than-{N}yquist modulation},
  author={Anderson, John B. and Rusek, Fredrik},
  booktitle={Proc. IEEE International Symposium on Information Theory (ISIT)},
  pages={2301--2304},
  year={2007}, month=jun,
  address={Nice, France}
}

@book{anderson2017bandwidth,
  title={Bandwidth Efficient Coding},
  author={Anderson, John B.}, 
  chapter={7},
  year={2017},
  publisher={John Wiley \& Sons}
}

@article{jaffal2022pulses,
  title={Pulses with minimum residual intersymbol interference for faster than {N}yquist signaling},
  author={Jaffal, Youssef and Alvarado, Alex},
  journal={IEEE Communications Letters},
  volume={26},
  number={11},
  pages={2670--2674},
  year={2022}, month=nov,
}

@inproceedings{jaffal2019achievable,
  title={Achievable rates using {PAM} time-limited pulses over band-limited channels: From {N}yquist to {FTN}},
  author={Jaffal, Youssef and Abou-Faycal, Ibrahim},
  booktitle={Proc. IEEE Wireless Communications and Networking Conference (WCNC)},
  pages={1--6},
  year={2019}, month=apr, address={Marrakech, Morocco},
}

@inproceedings{milojkovic2022pulseshaping,
  title={On pulse shaping for generalized faster than {N}yquist signaling with and without equalization},
  author={Milojkovi{\'c}, Jovan and Brki{\'c}, Sr{\dj}an and {\'C}erti{\'c}, Jelena},
  booktitle={Proc. International Conference on Electrical and Computer Engineering (IcETRAN)},
  pages={1--4}, month=jun,
  year={2022}, 
  address={Novi Pazar, Serbia}
}

@article{makarov2020optimizing,
  title={Optimizing the shape of faster-than-{N}yquist ({FTN}) signals with the constraint on energy concentration in the occupied frequency bandwidth},
  author={Makarov, Sergey B. and Liu, Mingxin and Ovsyannikova, Anna S. and Zavjalov, Sergey V. and Lavrenyuk, Ilya I. and Xue, Wei and Qi, Junwei},
  journal={IEEE Access},
  volume={8},
  pages={130082--130093},
  year={2020},
  publisher={IEEE}
}

@inproceedings{zhou2012generalized,
  title={Generalized faster-than-{N}yquist signaling},
  author={Zhou, Jing and Li, Daoben and Wang, Xuesong},
  booktitle={Proc. IEEE International Symposium on Information Theory (ISIT)},
  pages={1478--1482},
  year={2012}, month=jul, address={Cambridge, MA, USA},
}

@article{rusek2015ungerboeck,
  title={40 years with the {U}ngerboeck model: A look at its potentialities [lecture notes]},
  author={Rusek, Fredrik and Colavolpe, Giulio and Sundberg, Carl Erik W},
  journal={IEEE Signal Processing Magazine},
  volume={32},
  number={3},
  pages={156--161},
  year={2015}, month=may,
  publisher={IEEE}
}

@article{bedeer2017very,
  title={A very low complexity successive symbol-by-symbol sequence estimator for faster-than-{N}yquist signaling},
  author={Bedeer, Ebrahim and Ahmed, Mohamed Hossam and Yanikomeroglu, Halim},
  journal={IEEE Access},
  volume={5},
  pages={7414--7422},
  year={2017},
  publisher={IEEE}
}

@article{fan2018mlse,
  title={{MLSE} equalizer with channel shortening for faster-than-{N}yquist signaling},
  author={Fan, Jiancun and Ren, Yajie and Zhang, Ying and Luo, Xinmin},
  journal={IEEE Photonics Technology Letters},
  volume={30},
  number={9},
  pages={793--796},
  year={2018}, month=may,
  publisher={IEEE}
}

@article{li2020code,
  title={Code-based channel shortening for faster-than-{N}yquist signaling: {R}educed-complexity detection and code design},
  author={Li, Shuangyang and Yuan, Jinhong and Bai, Baoming and Benvenuto, Nevio},
  journal={IEEE Transactions on Communications},
  volume={68},
  number={7},
  pages={3996--4011},
  year={2020}, month=jul,
}

@article{takeshita2002frame,
  title={On the frame-error rate of concatenated turbo codes},
  author={Takeshita, Oscar Y and Collins, Oliver M and Massey, Peter C and Costello, Daniel J},
  journal={IEEE Transactions on Communications},
  volume={49},
  number={4},
  pages={602--608},
  year={2001}, month=apr,
  publisher={IEEE}
}

@article{dolinar1995weight,
  title={Weight distributions for turbo codes using random and nonrandom permutations},
  author={Dolinar, Sam and Divsalar, Dariush},
  journal={The Telecommunications and Data Acquisition Report 42-122},
  pages={56--65},
  year={1995}, month=aug,
}

@article{yang20256g,
  title={6{G}-oriented {LDPC}-coded faster-than-{N}yquist signaling: {C}ode design and performance analysis},
  author={Yang, Jiayi and Wang, Qianfan and Li, Shuangyang and Kang, Peng and Ma, Xiao and Bai, Baoming and Caire, Giuseppe and Wang, Xianbin},
  journal={IEEE Journal on Selected Areas in Communications},
  year={2025},
  doi={10.1109/JSAC.2025.3648707},
  note={{E}arly Access}
}

@article{li2026faster,
  title={Faster-than-{N}yquist signaling for next-generation wireless: {P}rinciples, applications, and challenges},
  author={Li, Shuangyang and Yuksel, Melda and Xu, Tongyang and Sugiura, Shinya and Yuan, Jinhong and Caire, Giuseppe and Hanzo, Lajos},
  journal={IEEE Communications Standards Magazine},
  year={2026},
  note={{E}arly Access},
  doi={10.1109/MCOMSTD.2025.3649943},
}

@article{verdu1994general,
  title={A general formula for channel capacity},
  author={Verd{\'u}, Sergio and Han, T. S.},
  journal={IEEE Transactions on Information Theory},
  volume={40},
  number={4},
  pages={1147--1157},
  year={1994}, month=jul,
  publisher={IEEE}
}

@article{turin1960characteristic,
  title={The characteristic function of {H}ermitian quadratic forms in complex normal variables},
  author={Turin, George L.},
  journal={Biometrika},
  volume={47},
  number={1/2},
  pages={199--201},
  year={1960}, month=jun,
  publisher={JSTOR}
}

@article{slepian1961prolate,
  title={Prolate spheroidal wave functions, {F}ourier analysis and uncertainty—{I}},
  author={Slepian, David and Pollak, Henry O.},
  journal={Bell System Technical Journal},
  volume={40},
  number={1},
  pages={43--63},
  year={1961}, month=jan,
  publisher={Wiley Online Library}
}
